\documentclass[]{article}
\usepackage[a4paper, total={7in, 10.5in}]{geometry}
\usepackage{bbm}
\usepackage{mathtools}
\usepackage{amsmath}
\usepackage{float}
\usepackage{natbib}
\usepackage{subcaption}
\usepackage{hyperref}
\usepackage{natbib} 

\title{Estimation of Optimal Dynamic Treatment Regimes using Gaussian Processes Emulation}
\author{Daniel Rodriguez Duque, David A. Stephens, Erica E.M. Stephens}

\begin{document}

\maketitle

\begin{abstract}
In precision medicine, identifying optimal sequences of decision rules, termed dynamic treatment regimes (DTRs), is an important undertaking. One approach investigators may take to infer about optimal DTRs is via Bayesian dynamic Marginal Structural Models (MSMs). These models represent the expected outcome under adherence to a DTR for DTRs in a family indexed by a parameter $ \psi $; the function mapping regimes in the family to the expected outcome under adherence to a DTR is known as the value function. Models that allow for the straightforward identification of an optimal DTR may lead to biased estimates. If such a model is computationally tractable, common wisdom says that a grid-search for the optimal DTR may obviate this difficulty. In a Bayesian context, computational difficulties may be compounded if a posterior mean must be calculated at each grid point. We seek to alleviate these inferential challenges by implementing Gaussian Process ($ \mathcal{GP} $) optimization methods for estimators for the causal effect of adherence to a specified DTR. We examine how to identify optimal DTRs in settings where the value function is multi-modal, which are often not addressed in the DTR literature. We conclude that a $ \mathcal{GP} $ modeling approach that acknowledges noise in the estimated response surface leads to improved results. Additionally, we find that a grid-search may not always yield a robust solution and that it is often less efficient than a $ \mathcal{GP} $ approach. We illustrate the use of the proposed methods by analyzing a clinical dataset with the aim of quantifying the effect of different patterns of HIV therapy.   
\end{abstract}

\section{Introduction}
In health research, as data capture and storage capacities improve, the questions researchers ask are becoming more complex. Ambitious questions may be posed in the quest for precision medicine where investigators seek to tailor treatment to patient-specific characteristics through stages of the clinical decision-making process. This tailoring requires sets of decision rules, termed dynamic treatment regimes (DTRs), that take patient information as inputs and that output a treatment recommendation at each stage of the treatment decision-making process. Often, researchers are interested in asking causal questions in relation to these DTRs. Most directly, such questions focus on quantifying what is the causal effect of adherence to a specific DTR and identifying what might the optimal DTR be. The search for an optimal therapy is an important one in medicine, as it aims to avoid over-treatment, all while providing sufficient care to arrive at the targeted outcome. Answering questions about DTRs is challenging, even in data-rich environments; more data may imply that we can ask more challenging questions, but the curse of dimensionality tells us that we cannot altogether escape thinking about statistical models. In this work, we examine how Gaussian Processes ($ \mathcal{GP} $s) may yield a strategy that allows for the robust identification of optimal DTRs. 

In the frequentist setting, inferential methods for DTRs have been traditionally performed via semi-parametric models. These include dynamic marginal structural models (MSMs) \citep{Orellana2010(a)}, g-estimation of structurally nested mean models \citep{Robins1986}, Q-learning \citep{Zhao2009} and outcome weighted learning \citep{zhao2012}.  For Bayesians, where modeling the entire probabilistic dynamics is often required for inference, a variety of methods for DTRs have also been proposed, including those of \cite{arjas2010optimal}, \cite{Saarela2015(a)}, \cite{xu2016}, \cite{murray2018}, and \cite{DRD}. Although much of the Bayesian literature in this area has focused on adapting existing frequentist estimation approaches for DTRs, the computationally intensive nature of Bayesian inference limits their usability. In this work, we focus on eliminating some of the modeling challenges with DTRs in order improve the usability of methods, be they frequentist or Bayesian. 

Our work is motivated by Dynamic MSMs, where a Bayesian version was recently proposed \citep{DRD}. These allow for the estimation of the value function of DTRs in a family $ \mathcal{G}$ indexed by a, possibly multi-dimensional, parameter $ \psi\in \mathcal{I} $. In a family of DTRs of the form ``treat when covariate value $ x $ exceed a threshold $ \psi $", researchers may posit a marginal mean model such as $E[Y^{g^\psi}]=\beta_0 + \beta_1 \psi + \beta_2 \psi^2$. Unfortunately, we cannot be certain that this model is correctly specified or that it is sufficiently flexible to correctly identify the optimal regime. One way around this issue is to estimate the expected outcome under each regime in the family, if there is a finite number of them, or to estimate the expected outcome of a large set of regimes and then extrapolate the value to other regimes. This essentially amounts to a grid-search and is an appealing approach as we have access to standard estimators for the expected outcome under adherence to a DTR. Unfortunately, this may be computationally intensive, particularly in settings with many stages, complex decision rules, and a variety of confounders. Computational challenge may be compounded in Bayesian settings where sampling of a posterior distribution is often required. Even if a grid-search is feasible, it has not been established in the literature whether it reliably identifies the optimal regime or whether there are other robust approaches that use data more efficiently.

The contribution of this work is to examine how to utilize computer experiments to identify optimal DTRs in a family $ \mathcal{G} $; an optimal DTR is one that maximizes the value function. In a DTR context, a “computer experiment" should sample the value function at strategically chosen points with the aim of approximating the value function, all while limiting the number of samples obtained. These experiments should begin by selecting an initial set of design points $ \mathcal{I} $, and then using an estimator for the value of a DTR at these points to arrive at a working model for the value function. This working model can then be utilized to select new points sequentially using a criterion, known as an infill criterion, that specifies where an optimum may be. We focus on the Expected Improvement criterion \cite{jones1998} which has been well studied and is known to balance exploration of the input space with exploitation of the optimizing region. Using this approach, we focus on methods that yield models more flexible than those used with Dynamic MSMs and that allow for the sequential sampling of additional points in order to improve the estimation of optimal DTRs. The models used are obtained via a $ \mathcal{GP} $ prior, with parameters fit using empirical Bayes or maximum a posteriori (MAP) inference. This is a novel approach to identifying optimal DTRs that has not previously been explored in the precision medicine literature. The computer experiment faces an additional challenge in that we do not have access to the value function, but rather to an estimator for the value of a DTR which can be evaluated point-wise. Via simulations, we find that a $ \mathcal{GP} $ modeling approach that acknowledges uncertainty in the estimated regime values can successfully identify optimal DTRs.  Additionally, we find that a grid-search for the optimum may not always be the best solution, especially in multi-modal settings which challenges the received wisdom that a grid-search is as reliable as other methods. We find that computer experiments via $ \mathcal{GP} $s can perform better than a grid-search, all while using fewer experimental points. In addition to these contributions, we illustrate how to use the discussed methods to perform a statistical analysis on clinical data arising from HIV therapy. 

\section{Background}
Traditionally, computer experiments to identify an optimum were performed using regression-based methods fit on a set of experimental points. However, these methods may not be well suited for identifying  optimal responses. \cite{huang2006global} mention that regression-based approaches may be inefficient as they attempt to predict the response curve over the entire feasible domain, as opposed to the neighborhood of the optimum. Additionally, regression models are usually relatively simple, and may not fit complex systems adequately over the entire domain similar to the issues we described with Dynamic MSMs. Consequently, more recent literature on computer experiments focuses on approaches using $ \mathcal{GP} $s and sequentially sampling new experimental points most relevant to the optimization. 

A $\mathcal{GP}$ is a stochastic process for which all outcome vectors, regardless of the dimension, have a multivariate Normal distribution. Models arising from the $\mathcal{GP}$ assumption are often termed kriging models. These models are widely used in two settings: 1) where researchers would like to fit a flexible model, which may be used for prediction in unobserved locations; or 2) where researchers are working with a function that is expensive to evaluate and would like to identify the optimum of this function, all while limiting the number of function evaluations. The latter falls into the domain of the computer experiments literature, which is most relevant to our work. In addition to the $ \mathcal{GP} $ assumption, much of this literature focuses on settings where the input-output relationship is known. Our setting is nuanced as we only have access to an estimated (noisy) output for any given input. Consequently, we must think carefully about the problem characteristics before developing an optimization strategy. 

\cite{sacks1989} were among the first researchers to explore using $ \mathcal{GP} $ for computer experiments. Later, \cite{currin1991} used a similar methodology but in a Bayesian context. \cite{o1998} argued that a Bayesian perspective is crucial for computer experiments with deterministic functions as, for a fixed input, the output does not change. Consequently, uncertainty about the response surface is not aleatory. Notwithstanding, a fully Bayesian treatment of this problem is often highly complex and some compromises must be made. We take this into consideration and seek practical methods that balance the benefits of the Bayesian and frequentist approach. One concern may be, as with any optimization procedure, that there exist local maxima within the operating domain of interest, making the identification of a global maximum more challenging. \cite{jones1998} emphasize that a computer experiments methodology  based on $ \mathcal{GP}s $ is good for modeling non-linear multi-modal functions. In addition to the $ \mathcal{GP} $ model, which requires specification of a covariance function an infill criterion must be specified. There are a variety of infill criteria in the literature; we make use of the Expected Improvement criterion although some care should be taken as it encounters theoretical problems in settings with noisy outputs.  We will examine these issues in what follows.  


\section{Problem Characteristics}

Before characterizing our specific inferential problem, let us fix some terminology regarding the surfaces of interest. The \textit{value surface} refers to the true relationship between a DTR, $ g^\psi $, and its value $ E[g^\psi] $; the \textit{estimation surface} refers to the surface obtained by point-wise evaluation of an estimator $ \hat{E}[g^\psi] $ for varying $ \psi\in \mathcal{I} $; note that $ \mathcal{I} $ is an index set that can be continuous or discrete. When we make use of an inverse probability weighted (IPW) estimator to obtain this surface, we refer to this estimation surface as the \textit{IPW-surface}. Lastly, the \textit{emulation surface} is the posterior mean of a given $ \mathcal{GP} $ of interest that is meant to approximate the \textit{value surface}.

Our setting is unique in that we are looking to emulate the value surface by only observing values from the estimation surface. As the estimation surface is produced by evaluating an estimator point-wise in a relevant domain, this function exhibits a non-smooth quality, and we are in a setting where the observed output is a noisy version of the true output. We will see that this non-smoothness may affect the results obtained via a grid-search. As investigators, we are likely interested in smoothing out this noisy surface, believing the true value function to be smooth; a $ \mathcal{GP} $ model allows for this possibility.

 The lack of smoothness of the estimated surface is mainly a consequence of using a finite sample size to estimate the value of regimes in $ \mathcal{G} $. Furthermore, we may ask whether adequately capturing this noise structure improves the resulting inference and whether this noise structure is homoskedastic or heteroskedastic There are several components in the data analysis that may lead to a heteroskedastic structure. These include 1) measurement error, possibly including more variability in treatment arm than in the response arm; 2) relatively smaller sample-size in some regions of the regime index set than others; and 3) patient responses being more distant from the value function in some areas of the index set than others. We more precisely illustrate these considerations in the following sections and in Appendix A. 

Kriging methods can also be used in settings with noisy observations; \cite{picheny2014noisy} provide an overview of these methods. Indeed, stochastic kriging is nuanced and not all methods are applicable in all settings. Stochastic kriging  is often utilized when emulating a response surface where at each experimental point the output varies when re-evaluated at the same input. In settings that do not involve sequential sampling of experimental points this definition is sufficient, as a model is fit on a fixed and known set of points. However, when sequential sampling is required, more care should be taken in defining the problem. There are some settings where we observe a noisy function but where there is no uncertainty in the output when re-evaluating at already sampled points. \cite{forrester2006design} explain that in this setting there is no uncertainty in the output, even if there is noise around the true curve. In other settings, re-evaluating at the same input yields varying outputs. This detail is consequential when identifying infill criteria for stochastic kriging. In some cases, we gain information by re-sampling at the same data-point --- in others we do not. Our motivating DTR setting relates most to the case where a curve exhibits a characteristic jitter but where there is no uncertainty in the output of already sampled points.

Stochastic kriging has focused on methods with homoskedastic noise; however there is a growing literature on incorporating heteroskedastic noise in the inferential procedure. For example, \cite{ankenman2008stochastic} and \cite{yin2011kriging} incorporate heteroskedastic noise by estimating the noise variance at design points; these authors' approach requires that the function of interest be evaluated at the design points multiple times. \cite{frazier2011value} also discuss heteroskedastic error and propose a method for financial time series.  A fully Bayesian approach is presented by \cite{goldberg1997regression} who seek to place a $ \mathcal{GP} $ prior on the log noise, yielding two $ \mathcal{GP} $ priors. Indeed a fully Bayesian treatment is computationally intensive, but some work has been done on alleviating these issues;  \cite{wang2014gaussian} has looked at fast MCMC procedures for $\mathcal{GP}$s with heteroskedastic noise. Thinking about practicality, \cite{kersting2007most} follow the same approach as \cite{goldberg1997regression}, however they focus on most likely heteroskedastic $\mathcal{GP}$s to estimate the input-dependent noise level. \cite{zhang2020improved} offer an improvement on most likely heteroskedastic $ \mathcal{GP} $s by providing an approximately unbiased estimator for the input-dependent noise. In what follows we will examine the performance of the latter approach.


From the above considerations, we regard a $\mathcal{GP}$ that acknowledges noise is an important component of the model; it remains to examine what criteria may be used to sample points sequentially. \cite{picheny2013benchmark} provide a review of infill criteria used for stochastic kriging. \cite{frazier2016bayesian} emphasize that the Expected Improvement criterion benefits from some optimality results in the deterministic setting but that these benefits are lost in noisy stochastic settings. In particular, in deterministic settings, the Expected Improvement criterion ensures the true optimum will be identified as the number of experimental points increases. This result hinges on the posterior variance at already sampled points being zero \citep{locatelli1997bayesian}, but this property is not necessarily present in stochastic settings. Many infill criteria allow for re-evaluations at already sampled points, but this is not desirable in our setting. There are other technical issues in revisiting experimental points with the  $ \mathcal{GP} $, for example ill-conditioned matrices.   \cite{forrester2006design} propose a solution for using the Expected Improvement in noisy settings by utilizing a re-interpolation approach for optimization. This is the approach that we explore.


\section{Methods}

We consider a sequential decision problem with $T$ decision points and a final outcome $Y$ to be observed at stage $T+1$. Decisions taken up to stage $t$ give rise to a sequence of treatments $\bar{z}_t=(z_1,...,z_t)$, $z_j \in \{0,1\}$. At each stage $t$, a set of covariates $x_t$ is available for decision-making and it is assumed that these consist of all time-fixed and time-varying confounders.  To denote covariate history up to time $t$, we write  $\bar{x}_t=\{x_1,...,x_t\}$. Subscripts are omitted when referencing history through stage $ T $. Then, all patient information is given by $ b=(\bar{x},\bar{z},y) $. We denote a DTR-enforced treatment history by $g(\bar{x})=\bar{g}(\bar{x})=(g_1(x_1),...,g_T(\bar{x}_T))$. Throughout, we will consider a family of DTRs, indexed by  $\psi \in \mathcal{I}$ to give $ \mathcal{G}=\{g^\psi(\bar{x});\psi\in \mathcal{I}\}$. In general, we allow $ \psi $ to be a $ p $-dimensional column vector. The index is omitted when it is clear that our focus lies on a single DTR. Based on these definitions, we posit that values $ upsilon_i $ on the estimation surface are a noisy realization of the value surface $ f(\psi) $ as given by the following relationship:
\begin{equation}
	\upsilon_i=f(\psi_i)+\epsilon_i \;, \; \epsilon_i \sim N(0,\gamma^2(\psi_i)), \; i=1,...,m.
	\label{eq:setup}
\end{equation}
Our target of inference is the value surface $ f $ for which there is epistemic uncertainty. As equation \ref{eq:setup} makes clear, this problem is further complicated as we do not observe $f$, but instead mereley a noisy version of it. To fix the notation about this model, suppose we have data $\mathcal{D}=\{\psi_i,\upsilon_i\}_{i=1}^m$. Then define the following vector quantities $ \psi=(\psi_1,...,\psi_m)^T $, $\upsilon=(\upsilon_1,...,\upsilon_m)$ and $ f=(f_1,...,f_m) $. We also define $ \bar{\gamma}^2=(\gamma^2(\psi_1),...,\gamma^2(\psi_m)) $. Note that these are observations taken on the estimation surface. We have control of the observations that we sample from this surface, and these contrast the observations on the sample $(\bar{x},\bar{z},y)$ which are fixed at a sample size $ n $.

To perform inference, we place a prior on $ f $, which represents our belief about the value function associated with a family of DTRs indexed by $ \psi \in \mathcal{I} $. We choose this to be a prior $d\pi(f)$ in a function space $ f \in \mathcal{F}$. Heuristically, as in \cite{shi2011gaussian}, updating can be done via the equation:
\begin{equation}
	P(f\in A|\mathcal{D},\gamma^2)=\int_A \frac{p(\upsilon|f,\gamma^2) d\pi(f) }{\int_\mathcal{F}p(\upsilon|f,\gamma^2)d\pi(f)}, \; \;  A \subset \mathcal{F}.
\end{equation}
More concretely, the prior that we make use of is a $ \mathcal{GP} $ prior, which has the consequence that for any finite set of observations $\psi $,  $f|\psi \sim N(\mu_{0f},K)$. $ K $ is a covariance matrix calculated via a covariance function $k(\psi_i,\psi_j)$ that is parameterized by parameters $ (\theta_f, \sigma^2_f) $, with $ \theta_f $ being a vector with entries $ \theta_{fd} $ controlling the correlation between points in the $ dth $ dimension and $ \sigma^2_f $ being a parameter that scales the correlation function to yield the covariance function.  The $ \mathcal{GP} $ requires specification of a set of hyperparameters $\eta_f=(\mu_{0f},  \theta_f, \sigma^2_f) $. Without further knowledge of the problem, it is challenging to specify values for these hyperparameters. Specifying priors for these hyperparameters is possible, but it may increase computational challenges to carry out a fully Bayesian treatment of this problem. More commonly, empirical Bayes is used to estimate the hyperameters via maximum likelihood, as in \cite{shi2011gaussian}. Alternatively, MAP estimation of the hyperparameters may be used. Conditional on fixing these hyperparameters, at their MAP or empirical Bayes estimates, standard arguments for the conditional distribution of a multivariate normal distribution yield the posterior distribution at a new point $ \psi_{m+1} $ to be:
\begin{align}
	\begin{split}
	f_{m+1}|&\psi_{m+1},\eta_f, \bar{\gamma}^2, \mathcal{D}  \sim  N(\mu_{f_{m+1}}, \sigma^2_{f_{m+1}})\\
	&\mu_{f_{m+1}}=\mu_{0f}+k_{m+1}^T(K+S)^{-1}(\upsilon-\mu_{0f})\\
	&\sigma^2_{f_{m+1}}=k(\psi_{m+1},\psi_{m+1})-k_{m+1}^T(K+S)^{-1}k_{m+1},
	\end{split}
\label{eq:regress1}
\end{align}
where $ S $ is a diagonal matrix of noise variances with $ii$th entry equal to $ \gamma^2_i =\gamma^2(\psi_i)$; $ k_{m+1}=(k(\psi_1,\psi_{m+1}),...,k(\psi_m,\psi_{m+1})) $ is the variance vector between already sampled points and the new point $ \psi_{m+1} $. In the empirical Bayes setting, the parameters are fixed values. Consequently, they need not be included in the conditioning, we do this however for compatibility with the MAP approach. Note that unlike the more well known $ \mathcal{GP} $ model for computer experiments, this model does not necessarily interpolate the observed data. That is, $ \mu_{f_{m+1}} $ does not necessarily perfectly predict the observed data points.  This is desirable, as we seek a smooth response curve, but we only have access to the noisy estimation surface.  To recover the interpolating model, we set $ \gamma^2(\psi_i)=0 \; \forall i $.  As \cite{forrester2006design} point out, the interpolation property of a $ \mathcal{GP}  $ occurs when there is no measurement error in the data observation mechanism and comes from noting that the posterior variance is zero at already sampled points. In what follows, we will more closely examine non-interpolating scenarios.
The remaining quantity of interest is the posterior distribution for the noisy observations: 
\begin{align}
	\begin{split}
		\upsilon_{m+1}|&\psi_{m+1},\eta_f ,\bar{\gamma}^2, \gamma^2_{m+1}, \mathcal{D}  \sim  N(\mu_{\upsilon_{m+1}}, \sigma^2_{v_{m+1}})\\
		&\mu_{\upsilon_{m+1}}=\mu_{f_{m+1}}\\
		&\sigma^2_{\upsilon_{m+1}}=k(\psi_{m+1}, \psi_{m+1}) - k_{m+1}^T(K+S)^{-1}k_{m+1} + \gamma^2_{m+1}.
	\end{split}
\label{eq:regress2}
\end{align}

\subsection{Homoskedastic Inference}
If noise is a concern, an interpolating $ \mathcal{GP} $ approach may not be adequate, and we may look to allow for noise around the surface. If we assume that the noise variance is homoskedastic, then we have that  $\gamma^2(\psi_i)=\gamma^2\; \forall i$.  Under an empirical Bayes approach our posterior of interest is $ p(\upsilon_{m+1}|\psi_{m+1}, \mathcal{D})= p(\upsilon_{m+1}|\psi_{m+1},\eta_f ,\gamma^2, \mathcal{D}  )$. To compute  values for the hyperparameters, we maximize $ p(\upsilon|\psi,\eta_f,\gamma^2) $. Efficient computational approaches to identifying the maximizers of this marginal likelihood can be found in \cite{park2001efficient} and \cite{roustant2012}. With access to this model, we could additionally combine it with MAP estimation of $ \theta_f $ in order arrive at an approximation for $p(\upsilon_{m+1}|\psi_{m+1}, \mathcal{D})$. This requires maximizing $ p(\eta_f, \gamma^2|\mathcal{D}) $ with respect to $ \eta_f,\gamma^2 $ in order to obtain $ \eta_f^{map}, \gamma^{2,map} $. MAP estimation then uses the approximation $ p(\eta_f,\gamma^2|\mathcal{D})\approx \mathbbm{1}_{(\eta_f^{map},\gamma^{2,map})}(\eta_f,\gamma^{2}) d(\eta_f,\gamma^2)$ in order to arrive at the posterior predictive distribution as: 
\begin{equation}
p(\upsilon_{m+1}|\psi_{m+1}, \mathcal{D}) \approx \int p(\upsilon_{m+1}|\psi_{m+1},\eta_f,\mathcal{D}) \mathbbm{1}_{(\eta_f^{map},\gamma^{2,map})}(\eta_f,\gamma^{2}) d(\eta_f,\gamma^2) =p(\upsilon_{n+1}|\psi_{m+1}, \eta^{map}_f,\gamma^{2,map},\mathcal{D}).\\
\end{equation}
 \cite{Lizotte2008} has examined MAP inference for deterministic computer experiments under a Log-Normal prior for $ \theta_f $.
 
\subsection{Heteroskedastic Inference}

Alternatively, we may believe that the response surface exhibits heteroskedastic noise. This poses special challenges as it requires performing inference for each  of the noise variances, $ \gamma_i $, in the observed data. For this, we examine an approach proposed by \cite{zhang2020improved} that places a second $ \mathcal{GP} $ prior on the regression residuals $e_i= |r_i|^q=|\upsilon_i - \mu_{\upsilon_i}|^q$, $ q\in \mathcal{Z}^+ $, with covariance function $ k_e(\psi_i,\psi_j) $ and parameters $\eta_e =(\mu_{0e}, \eta_e, \sigma^2_e)$. Authors show that under these assumptions a method of moments estimator for the input-specific noise variances can be arrived at via:
\begin{equation}
	E[|r_i|^q]=\frac{\gamma_i}{s(q)}, 
	\label{newway}
\end{equation}
where $ s(q) $ is a correction factor. When $ q=1 $, the estimator for the input-dependent noise is approximately $ \tilde{\gamma}_i=\sqrt{\pi/2}E[|r_i|]=\sqrt{\pi/2}\mu_{e_i} $, where $  \mu_{e_i} $ is the posterior mean of the second $ \mathcal{GP}$. 
A fully Bayesian computation that acknowledges uncertainty in $ \gamma_i $ would require an integral like:
\begin{equation}
	p(\upsilon_{m+1}|\psi_{m+1}, \eta_f,  \mathcal{D})=\int \int p_1(\upsilon_{m+1}|\psi_{n+1}, \eta_f, \bar{\gamma}^2, \gamma^2_{m+1}, \eta_e,\mathcal{D})p_2(\bar{\gamma}^2,\gamma_{m+1}^2|\psi_{n+1},\mathcal{D})d\bar{\gamma} d\gamma_{m+1}.
	\label{to_integrate}
\end{equation}
For known $ \bar{\gamma},\gamma_{m+1} $, sampling from $p_1$ is Normal with posterior mean and variance as described in equation (\ref{eq:regress2}). However, this computation is challenging because the $ \gamma^2 $ are unobserved. \cite{goldberg1997regression} provide an MCMC approach to allow for sampling from $ p_2 $ which computes the integral of interest, however this is computationally intensive. \cite{kersting2007most} proposed that $ p_2 $ be approximated by the most likely noise level. The most likely noise level is calculated as the posterior mean of a $ \mathcal{GP} $ that has been placed on $ \log(\gamma_i) $; recall that at each point the $ \mathcal{GP} $ is Normally distributed, therefore making the most likely value the $ \mathcal{GP} $ mean. \cite{zhang2020improved} provide an improved way to estimate $ \gamma_i $, as described in equation \ref{newway}, in order to yield the approximation $  \upsilon_{m+1}|\psi_{m+1}, \eta_f, \bar{\tilde{\gamma}}^2, \tilde{\gamma}_{m+1}^2, \mathcal{D} \sim N(\mu_{\upsilon_{m+1}},\sigma^2_{\nu_{m+1}}) $.  As in the empirical Bayes approach, $ \tilde{\gamma}^2_i $ are assumed known in the computation. Consequently, we can treat this posterior distribution as a $ \mathcal{GP} $ and perform inference as before.

In the following, we examine how to pair the homoskedastic and heteroskedastic models with the expected improvement criterion in order to arrive at a sequential sampling scheme. 
\subsection{Infill Criterion}
We return to the question of an appropriate infill criterion when we are interested in performing minimization. The Expected Improvement in our setting is given by: $EI(\psi)=E\left[\max(\upsilon(\psi)-\upsilon_{max})^{+}|\mathcal{D}\right]$. The expectation is taken with respect to the posterior distribution and $ \upsilon_{max}=max(\upsilon_1,...,\upsilon_m) $. Further computation yields:
\begin{equation*}
 EI(\psi)=(\mu_{\upsilon_{m+1}}(\psi)-\upsilon_{max})\Phi(\frac{\mu_{\upsilon_{m+1}}(\psi)-\upsilon_{max}}{\sigma_{\upsilon_{m+1}}(\psi)}) + \sigma_{\upsilon_{m+1}}(\psi) \phi(\frac{\mu_{\upsilon_{m+1}}(\psi)-\upsilon_{max}}{\sigma_{\upsilon_{m+1}}(\psi)})
\end{equation*}
  when $\sigma_{\upsilon_{m+1}}(\psi)>0 $ and $ 0 $ otherwise. $\Phi$ is the CDF of the Standard Normal distribution and $\phi$ is the corresponding pdf. 

\subsection{Re-interpolation}
As discussed, using the Expected Improvement as a criterion for sequential sampling may not be theoretically justified in a deterministic computer experiment with noisy observations, in particular when a regressive model is used rather than an interpolating model. Regressive models are ones that do not interpolate the sample data, like the homoskedastic and heteroskedastic models discussed above. The challenge in using the Expected Improvement with these models arises from the fact that the error $ \sigma_{\upsilon_{m+1}}(\psi) $ at sample points will be non-zero even though the output will not vary when the estimation function is re-evaluated at these sample points.  Consequently, convergence toward global optimum cannot be guaranteed \citep{locatelli1997bayesian}. \cite{forrester2006design} introduce a re-interpolation method that attains zero error at the sample locations. This can be done by building an interpolating $ \mathcal{GP} $ on the values predicted by the regressive model mean $ \mu_{\upsilon_{m+1}} $ and sequentially sampling using the Expected Improvement based on this mode. The procedure is termed re-interpolation because the interpolating model is build on the predicted mean values of the regressive model.
  
First, the re-interpolating procedure uses predictions at sample points obtained from the mean of  $\upsilon_{m+1}|\psi_{m+1} ,\mathcal{D}$ in order to create a new dataset $ \mathcal{D}' $. At sample point $ i $, we define the predicted values as $ \hat{\upsilon}_i=\mu_{\upsilon_{m+1}}(\psi_i) $ to yield responses $  (\hat{\upsilon}_1,...,\hat{\upsilon}_m) $ and new data $ \mathcal{D}'=\{\psi_i,\hat{\upsilon}_i\}_{i=1}^m$. Then using an interpolating $ \mathcal{GP} $ assumption on these data, we obtain a similar heuristic as before: 
$p(\hat{\upsilon}_{m+1}|\psi_{m+1}, \mathcal{D}') = p(\hat{\upsilon}_{m+1}|\psi_{m+1},\eta_{\hat{f}},\mathcal{D}')$, where the posterior mean and variance are given by: 
\begin{align*}
	\mu_{\hat{\upsilon}_{m+1}}=\mu_{0\hat{\upsilon}}+k_{m+1}^TK^{-1}(\hat{\upsilon}-\mu_{0\hat{\upsilon}})\\
	\sigma^2_{\hat{\upsilon}_{m+1}}=k(\psi_{m+1},\psi_{m+1}) - k_{m+1}^TK^{-1}k_{m+1},
\end{align*}
with $ \mu_{0\hat{v}} $ being the prior mean of the interpolating process. This re-interpolating procedure leads to two essential properties: 1) the posterior mean of the $ \upsilon $ and $ \hat{\upsilon} $ processes are the same i.e.
$ \mu_{\upsilon_{m+1}}=\mu_{\hat{\upsilon}_{m+1}}  $, and 2) the variance of the $ \hat{\upsilon} $ process is zero at already sampled points. The latter is the crucial characteristic required to preserve the optimality of the Expected Improvement criterion. With this re-interpolating model, the Expected Improvement can be calculated to determine new sampling locations. In Appendix B, we provide additional details on the equality of the two posterior means. \cite{forrester2006design} mention that the covariance  function $ K $ remains unchanged, so $ \eta_f $ does not need to be re-optimized. 

%
%
%
%
%

\subsection{Design of Experiments}

One component of the design of experiments is to determine the initial number of design points. \cite{loeppky2009choosing} investigate this issue and conclude that ten points per dimension is a reasonable rule-of-thumb when the dimension is less than five. we simply select them in equally spaced increments. Another option, for example, is to select design points randomly, but given the nature of our experiment, we aim to eliminate variability due to the initial sampling strategy. 

Another design element that must be considered is the covariance function. Some covariance functions in the $ \mathcal{GP} $ lead to smoother surfaces than others. One common choice of covariance is the $Mat\acute{e}rn$ covariance family. Common choices in this family are the $Mat\acute{e}rn_{5/2}$ covariance which is twice differential and the $Mat\acute{e}rn_{3/2}$ covariance which is differentiable once. These are examples of isotropic covariance functions, meaning that the correlation between points depends only on the distance between them. We focus on the  $Mat\acute{e}rn_{5/2}$ covariance given by:
\begin{equation}
	k(\psi_i,\psi_j)=\sigma^2_{f}\prod_{d=1}^{D}\left((1+\frac{\sqrt{5}|\psi_{id}-\psi_{jd}|}{\theta_{fd}} +\frac{5(\psi_{id}-\psi_{jd})^2}{3\theta_{fd}^2})\exp(\frac{-\sqrt{5}|\psi_{id}-\psi_{jd}|}{\theta_{fd}})\right),
\end{equation}
where $ D $ is the number of dimensions in the index vector.

\subsection{Estimation Surface}

As previously mentioned, the estimation surface can be produced with any estimator for the value of a DTR. In this work, we make use of the normalized IPW estimator. Then, for a family of interest, the estimator can be evaluated on a grid of $ \psi $s in order to yield the resulting estimation surface.  The normalized IPW estimator is given by 
\begin{equation}
 \frac{\sum_{i} w^{\psi}_iy_i}{\sum_{i} w^{\psi}_i}, \text{where} \; w^{\psi}_i=\frac{\mathbbm{1}_{\bar{g}^\psi(\bar{x})}(\bar{z})  }{\prod_{j=1}^{T}p(z_j |\bar{z}_{j-1},\bar{x}_{j})}.
\end{equation}
An additional layer of complexity is encountered if we are interested in using a Bayesian estimator to perform inference. This is because computing a posterior mean often requires sampling from the posterior distribution, which may be a computationally intensive task. In this case, a grid-search for the optimal DTR may become intractable. \cite{DRD} provide a Bayesian estimator for the value of a DTR by making use of inverse weighting and the Bayesian bootstrap. 

Generally, bootstrapping can allow for the quantification of sampling uncertainty. For example, in our setting, it may be that we are interested in quantifying uncertainty around the estimation surface. For a set of observations $(b_1,...,b_n)$, the bootstrap procedure samples each observations independently with replacement and with equal probability $ 1/n $ in order to estimate the quantity of interest. A similar procedure can be arrived at through a Bayesian lens as first proposed by \cite{Rubin1981}. This requires a posterior distribution that places a random probability $ \pi_i $ of sampling observation $ b_i $ in a bootstrapped sample; these probabilities have mean $ 1/n $ thereby connecting the procedure to frequentist bootstrap. This Bayesian bootstrap procedure can be arrived at by placing Dirichlet Process $ \mathcal{DP}(\alpha,G) $ prior on the data-generating distribution, where $ \alpha $ is a concentration parameter and where $ G $ is a base distribution. In particular, when $ \alpha $ is chosen such that $|\alpha| \to 0$, we obtain the Bayesian bootstrap as the posterior predictive distribution.  Under this specification, one sample drawn from the posterior $\mathcal{DP}$ is given by $p(b_{n+1}|\bar{b},\pi)=\sum_{i=1}^n \pi_i \mathbbm{1}_{b_i}(b_{n+1})$, where $\pi=(\pi_1,...,\pi_n) \sim Dir(1,...,1)$ is a sample from the Dirichlet distribution with all concentration parameters equal to one. Under these assumptions, any distribution sampled from the posterior  $ \mathcal{DP} $ is uniquely determined by $\pi$. For example the Bayesian bootstrap can be operationalized to quantify posterior belief about the population mean $ E[b_{n+1}|\bar{b}]=E_\pi[E[b_{n+1}|\bar{b},\pi]]$  by sampling weights $(\pi_1,...,\pi_n)$ and computing
\begin{equation}
 E[b_{n+1}|\bar{b},\pi]=\int_{b_{n+1}}b_{n+1} \sum_{i=1}^n \pi_i \mathbbm{1}_{b_i}(b_{n+1}) db_{n+1}=\sum_{i=1}^n \pi_i b_i.
\end{equation}
This quantity can be computed over many draws of the weights in order to obtain the full posterior distribution for the mean. Taking the mean across all these bootstrap samples results in an estimate for $  E[b_{n+1}|\bar{b}] $.

\subsubsection{Sources of Variation}
As we have already discussed, the estimation surface exhibits non-smoothness. In this section, we examine some possible sources of heteroskedastic variation. These considerations are most consequential for finite sample sizes. In this exploration, we limit ourselves to regimes of the form ``treat if $ x>\psi $", as this is a common regime in the literature, and it leads to clear examples about how heteroskedasticity is manifested. Additionally, we focused on the normalized IPW estimator for the value of a regime which uses only patients observed to adhere to the regime of interest. 

Note that in contrast to static treatment regimes, an individual can be simultaneously adherent with many DTRs \citep{Cain2010}. Furthermore, for a given sample with binary treatment, there are two response curves: the treated curve and the untreated curve. For a fixed $ \psi $, we can use the sample to estimate $ E[Y^{g^\psi}] $. Furthermore, for an increase  in $ \psi $ from $ \psi_1 $ to $ \psi_2 $, only treated patients can become non-adherent and only untreated patients can become adherent because of the form of rule under consideration. Only patients with covariate values $ \psi_1 \le x \le \psi_2 $ are eligible to become adherent/non-adherent. These properties are important in examining the variability in the estimation surface.

The first case we consider is heteroskedasticity due to distance from the value surface. This relates to how close/far the estimated treated and untreated curves are from the value surface. Recall that the value surface represents a population average; individual responses can vary substantially around this surface. For an increase in $ \psi $ from $ \psi_1 $ to $ \psi_2 $, there will be a set of patients who become non-adherent with regime $ \psi_2 $ and a set who become adherent. As the IPW estimator uses only observations on those adherent to a regime, if either the newly adherent/non-adherent patients have a response value that is far from the IPW-surface, then these observations will have a considerable influence on the estimate, especially for relatively small sample sizes. If the observations tend to have a response that is near the population average, then the IPW-surface will be less influenced by these observations. 

The second case is heteroskedasticity due to the noise structure at the individual level.  Consider an additive error term in the data-generating mechanism, such as: $  z\epsilon_1 + (1-z)\epsilon_2 $, where $\epsilon_1=N(0,5) $, $ \epsilon_2=N(0,0.5) $. We might not think this is an issue, as for estimation via an estimating equation, it does not matter whether noise is heteroskedastic or homoskedastic, so long as it has zero mean. However, when estimating the value surface for the purposes of identifying a maximum, this may be consequential. As $ \psi $ increases, we lose treated patients, and we gain untreated patients. This means that we lose observations with high variability and gain observations with low variability; this noise structure at the individual level transforms into heteroskedasticity at the estimator level. Now, we may ask when this data-generating mechanism may arise. One case may be when treatment leads to relatively reliable improvements, but lack of treatment  leads to disease progression taking on a variety of forms, and therefore leading to higher variability. 

The third consideration that may lead to heteroskedasticity in the estimation surface is the result of differing effective samples sizes across values of $ \psi $. It is well known that the IPW estimator for a regime $ \psi $ only uses patients who are adherent to the regime. Consequently, different regimes will use different number of patients to compute the value of the corresponding regime. This means that the estimator will exhibit differing levels of variability for a range of $ \psi $s.  In Appendix A, we further illustrate all three cases discussed.

\section{Simulations}
In what follows, we examine several data-generating mechanisms and DTRs to assess whether the $ \mathcal{GP} $ approaches presented do allow for the identification of optimal DTRs; we additionally compare these to a grid-search. We refer to the interpolating, homoskedastic, and heteroskedastic $ \mathcal{GP} $s as Int$ \mathcal{GP} $, HM$ \mathcal{GP} $, and HE$ \mathcal{GP} $, respectively. We present results for a sample size of $n=500$ with a $Mat\acute{e}rn_{5/2}$ covariance function. To produce the estimation surface, we make use of the normalized IPW estimator. To compare across modeling strategies, each analysis was performed on 500 Monte Carlo replicates. Appendices C through E examine scenarios with a sample size of $ n=1000 $, a $Mat\acute{e}rn_{3/2}$ covariance, and a Log-Normal prior.

\subsection{Simulation I}

\begin{figure}[H]
	\centering
	\begin{subfigure}[b]{0.4\linewidth} 
		\includegraphics[scale=0.3]{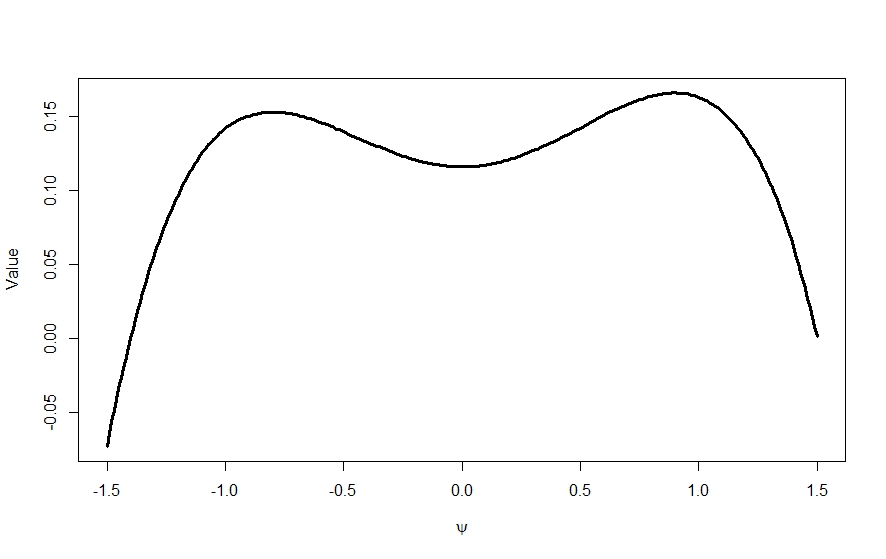}
		\subcaption{}
	\end{subfigure}
	\begin{subfigure}[b]{0.4\linewidth} 
		\includegraphics[scale=0.3]{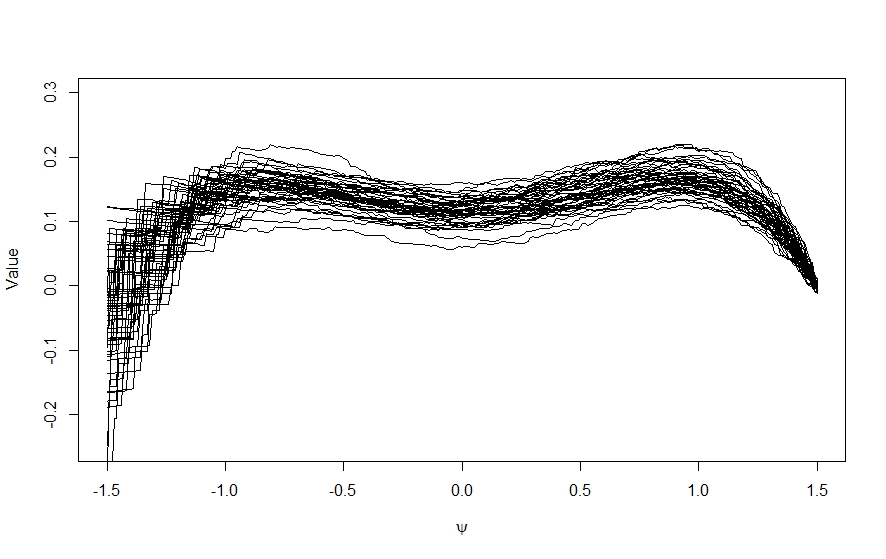}
		\subcaption{}
	\end{subfigure}
	\caption{Simulation I (a) Value function  (b) IPW estimates of value function, across 50 replicates.}
	\label{simulI_value} 
\end{figure}

 For this simulation, we generate covariate $x\sim U(-1.5,1.5)$, treatment $ z \sim Binom(p= expit(2x)) $, error distributions 	$ \epsilon_1=N(0,\sigma=0.25) $ $ \epsilon_2=N(0,\sigma=0.05) $ and final outcome  $ y=-(x+.8)x(x-.9)z+z\epsilon_1 + (1-z)\epsilon_2 $. We explore the regime ``treat if $x>\psi", \; \psi \in (-1.5,1.5) $. Note that $ expit(\cdot) $ refers to the inverse $ logit $ function. With this data-generating mechanism, the systematic component of $ y $ varies from -2 to 2.5 and the optimal regime represents a 5 \% improvement (in the range of $ y$) over the worst regime in the class.  In Figure \ref{simulI_value}, we observe the value function for this problem and the IPW-surface, across multiple replicates. It is visually evident that the function has two local maxima but only one global maximum at  $ \psi=0.9 $. There appears to be more variability for low values of $ \psi $ than for high values. Contrary to standard practice, a grid-search for the optimum may not work well, as evidenced by the large interquartile range (IQR) in Table \ref{simulI_grid_median}. 

For this simulation, the computer experiment was designed such that we sampled an initial set of design points in increments of 0.25, yielding an initial set of 13  points. Then, additional points were sampled sequentially using the Expected Improvement criterion, up to 25 additional points. All measures of variation correspond to Monte Carlo variation across replicates. We do not compute coverage probabilities for each $ \mathcal{GP} $, as for a fixed replicate, the uncertainty represented by the $ \mathcal{GP} $ is constrained to uncertainty in the IPW-surface resulting for a specific sample of size $ n $; it does not incorporate sampling uncertainty. Incorporating sampling uncertainty requires a more computationally intensive procedure, one that we explore in the case study.

\begin{figure}[H]
	\centering
		\hspace{-1.3cm}
	\begin{subfigure}[b]{0.3\linewidth} 
		\includegraphics[scale=0.2]{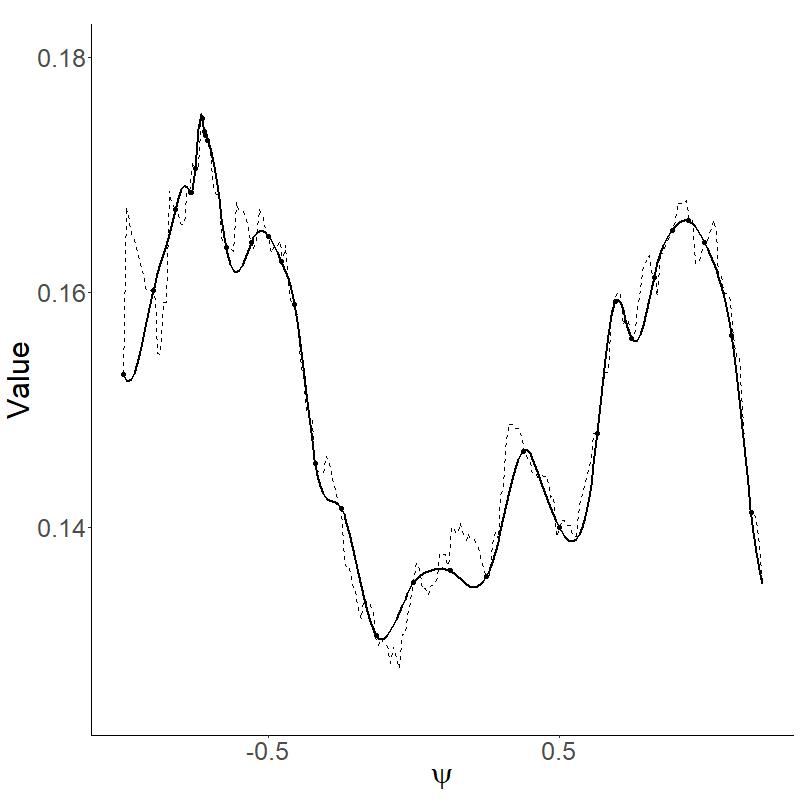}
		\caption{}
	\end{subfigure}
	\begin{subfigure}[b]{0.3\linewidth} 
		\includegraphics[scale=0.2]{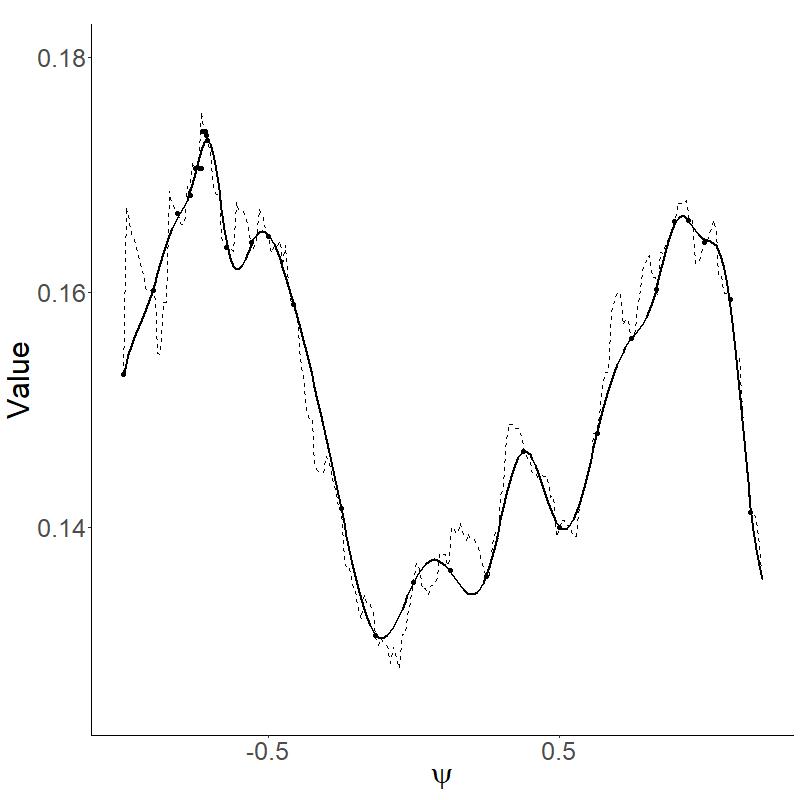}
				\caption{}
	\end{subfigure}
	\begin{subfigure}[b]{0.3\linewidth} 
		\includegraphics[scale=0.2]{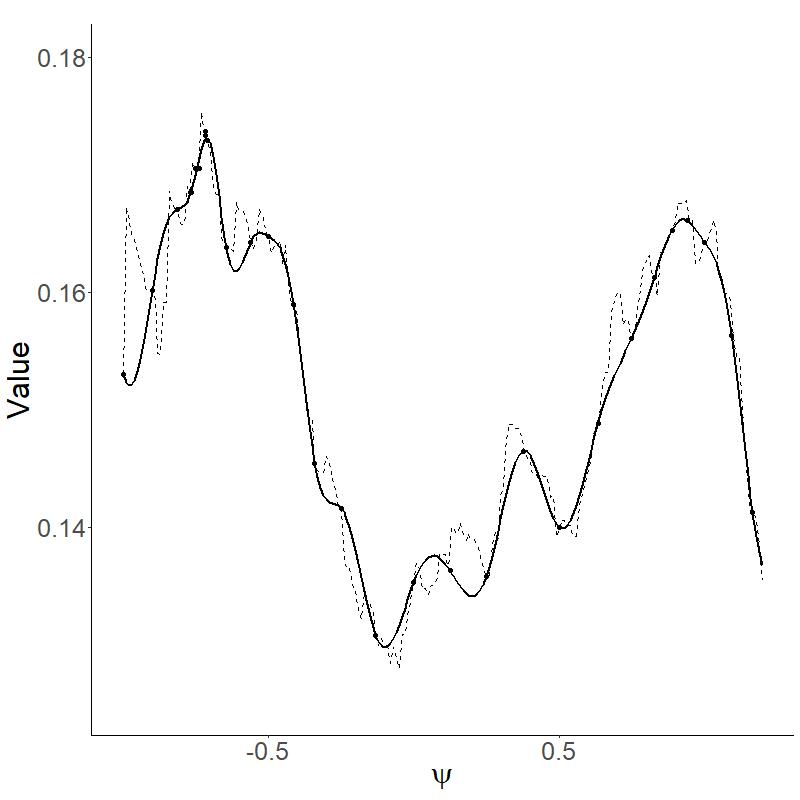}
				\caption{}
	\end{subfigure}
	\caption{Simulation I: Emulation surfaces at +25 points overlaid over the IPW-surface in restricted domain for $ \psi \in [-1,1.2] $ (a) Int$ \mathcal{GP} $ (b) HM$ \mathcal{GP} $ (c) HE$ \mathcal{GP} $.} 
	\label{simulI_paths_cov3_n500_prior0}
\end{figure}

\begin{table}[H]
	\centering
	\caption{ Results for grid-search with increments of 0.01 and $ n=500 $.  True $ \psi_{opt} $=0.9; true value at optimum  0.165.} 
	\begin{tabular}{rlll}
		\hline
		Statistic 		  & $ \hat{\psi}_{opt}$      & Value at $ \hat{\psi}_{opt}$  \\ 
		\hline
		 Mean (SD)      &  0.427  (0.800)  & 0.172  (0.022) \\ 
		Median (IQR)   &  0.860  (1.600)   & 0.171  (0.029) \\ 
		\hline
	\end{tabular}
	\label{simulI_grid_median}
\end{table}
From Figure \ref{simulI_paths_cov3_n500_prior0}, we see the results of the three modeling strategies for one replicate. These curves represent the posterior mean after sampling 25 additional points using the Expected Improvement as the infill criterion. In the figure, we restrict the domain of $ \psi $ for better visualization around the local and global optima, but in Appendix C the curves can be visualized for the entire decision space. From the figure, we see why the interpolating model is likely to under-perform; occasionally, due to noise in the fit, there will be a maximizer of the IPW-surface that is not close to 0.9. In these scenarios, the Int$ \mathcal{GP} $ will interpolate the data, whereas the other two methods can adjust the estimate based on the identified noise level. Careful examination of the graphs reveals that the interpolation is most consequential around the local optimizer $ \psi=0.8 $. Although HE$ \mathcal{GP} $ may assign higher variability to certain regions, it may also assign lower variability and become closer to interpolating. These plots contrast the differences between these methods, but they do not inform us about what will happen across many analyses. Consequently, we now look to assess their performance across multiple replications. Recall that unlike the context encountered in conventional computer experiments,here we have a target surface, the true value function, that for a given sample can be approximated by the IPW-surface. The IPW-surface is only an intermediary in the whole process, and we are interested in comparing the target surface with the emulation surface, in particular with respect to the optimizer.

Table \ref{simulI_tab_psi1_cov3_n500_prior0} shows the simulations results pertaining to the optimal threshold, for each modeling type. From this table, we note that the mean across replicates is not unbiased; this is due to the multi-modality of the problem. The variability is  higher for the grid-search and for the interpolating method than for the other methods. In what follows, due to the nature of the problem, we focus mainly on medians and interquartile ranges, though additional tables relating to means can be found in Appendix C.  We note further that the performance of the interpolating model degrades slightly as more samples are added, specifically with regard to the precision. The median obtained by the HM$ \mathcal{GP} $ is closest to the truth, and performance seems to increase slightly as more samples are added. For this simulation, all methods perform relatively well, even after few points are sampled. We note that at 25 additional samples, all three methods outperform the grid-search, which used 300 function evaluations, as measured by the median and IQR.

\begin{figure}[H]
	\centering
	\hspace{-1.3cm}
	\begin{subfigure}[b]{0.5\linewidth} 
		\includegraphics[scale=0.25]{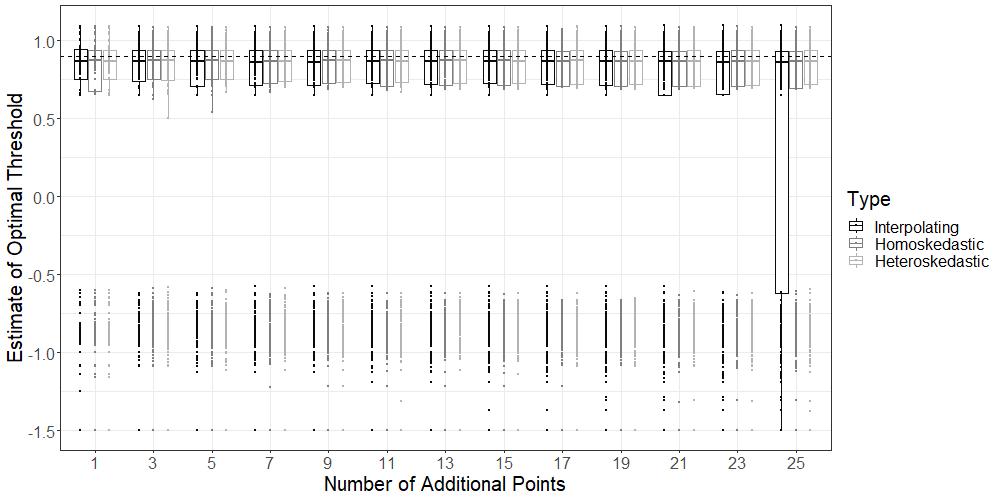}
	\end{subfigure}
	\begin{subfigure}[b]{0.5\linewidth} 
		\includegraphics[scale=0.25]{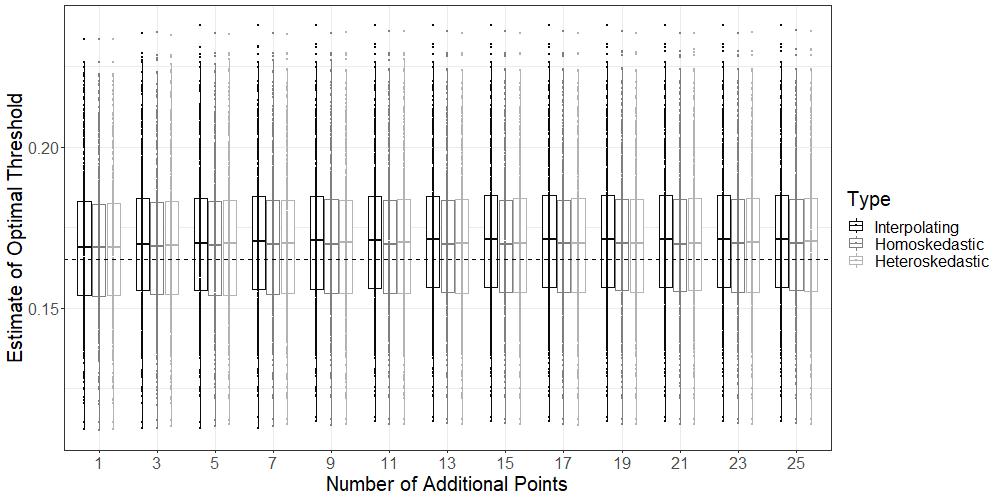}
	\end{subfigure}
	\caption{Simulation I: Boxplot at $ +m $ points; $ n=500 $ with 16 design points (a) Optimal $ \psi_1 $  (b) Value at optimum.} 
	\label{simulI_boxplot_cov3_n500_prior0}
\end{figure}

Table \ref{simulI_tab_medianvalue_cov3_n500_prior0} shows the consequences of the estimation procedure on the value of the optimal regime. We see that, like the grid-search values in Table \ref{simulI_grid_median}, these does not deviate as much as the optimizer. This is because the local and global optimizers in the value function have similar values.  Figure \ref{simulI_boxplot_cov3_n500_prior0} depicts the results for both the optimal threshold and for the value at the optimum; in panel (a) we see that the interpolating method appears to display worse performance as more points are sampled; this is an artifact of the interpolation that the method performs. From this simulation, we conclude that the HM$\mathcal{GP}$ and HE$\mathcal{GP}$, which acknowledge noise in the IPW-surface, yield results that are closest to the truth across replicates. We also conclude that any of the $ \mathcal{GP} $ modeling approaches outperform the grid-search, which additionally is less computationally efficient. In Appendix C, we find that for a larger sample size of $ n=1000 $ the performance of the grid-search improves to become comparable with the $ \mathcal{GP} $ approaches.

\begin{table}[H]
	\centering
	\caption{Simulation I: Optimal $\psi$ after $+m$ points; $ n=500 $ with 13 design points. True $ \psi_{opt} $= 0.9.} 
	\begin{tabular}{rlllllll}
		\hline
		Measure& & +1 & +5 & +10 & +15 & +20 & +25 \\ 
		\hline
		Mean SD    & Int$ \mathcal{GP} $ & 0.472 (0.760) & 0.474 (0.759) & 0.481 (0.755) & 0.475 (0.764) & 0.454 (0.779) & 0.440 (0.787) \\ 
		Mean SD    & HM$ \mathcal{GP} $  & 0.466 (0.766) & 0.501 (0.737) & 0.484 (0.751) & 0.477 (0.754) & 0.471 (0.757) & 0.469 (0.761) \\
		Mean SD    & HE$ \mathcal{GP} $  & 0.487 (0.751) & 0.504 (0.736) & 0.499 (0.741) & 0.479 (0.753) & 0.472 (0.759) & 0.476 (0.759) \\ 
		\hline
		Median IQR & Int$ \mathcal{GP} $ & 0.863 (0.194) & 0.867 (0.231) & 0.866 (0.208) & 0.867 (0.210) & 0.865 (0.240) & 0.861 (1.552) \\ 
		Median IQR & HM$ \mathcal{GP} $  & 0.874 (0.260) & 0.873 (0.189) & 0.871 (0.218) & 0.869 (0.226) & 0.866 (0.227) & 0.868 (0.237) \\ 
		Median IQR & HE$ \mathcal{GP} $  & 0.869 (0.186) & 0.868 (0.188) & 0.872 (0.206) & 0.866 (0.212) & 0.865 (0.219) & 0.865 (0.213) \\ 
		\hline
	\end{tabular}
	\label{simulI_tab_psi1_cov3_n500_prior0}
\end{table}

\begin{table}[H]
	\centering
		\caption{Simulation I: Value at optimum after $+m$ points, median (IQR); n=500 with 13 design points. True value at optimum: 0.165.} 
	\begin{tabular}{rllllll}
		\hline
		& +1 & +5 & +10 & +15 & +20 & +25 \\ 
		\hline
		Int$ \mathcal{GP} $  & 0.169 (0.029) & 0.170 (0.029) & 0.171 (0.029) & 0.171 (0.029) & 0.171 (0.029) & 0.171 (0.028) \\ 
		HM$ \mathcal{GP} $   & 0.169 (0.029) & 0.170 (0.029) & 0.170 (0.029) & 0.170 (0.029) & 0.170 (0.028) & 0.170 (0.028) \\ 
		HE$ \mathcal{GP} $   & 0.169 (0.029) & 0.170 (0.029) & 0.170 (0.029) & 0.170 (0.029) & 0.170 (0.029) & 0.171 (0.029) \\ 
		\hline
	\end{tabular}
	\label{simulI_tab_medianvalue_cov3_n500_prior0}
\end{table}

\subsection{Simulation II}

\begin{figure}[H]
	\centering
	\begin{subfigure}[b]{0.45\linewidth} 
		\includegraphics[scale=0.28]{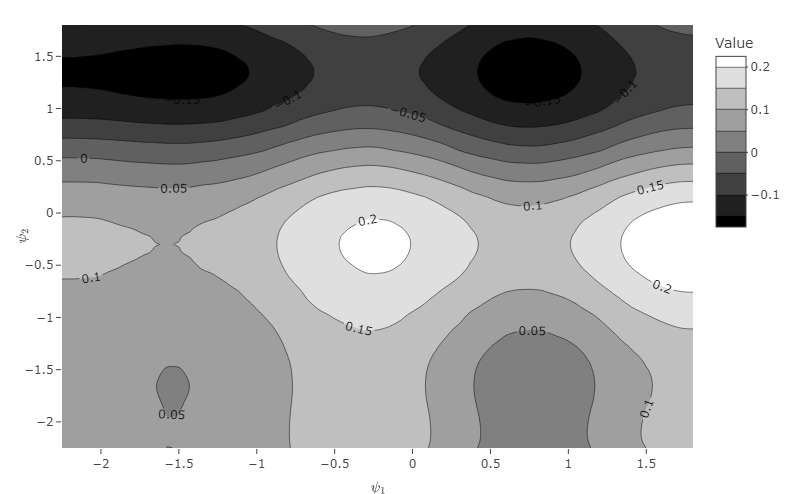}
		\subcaption{}
	\end{subfigure}
	\begin{subfigure}[b]{0.45\linewidth} 
		\includegraphics[scale=0.28]{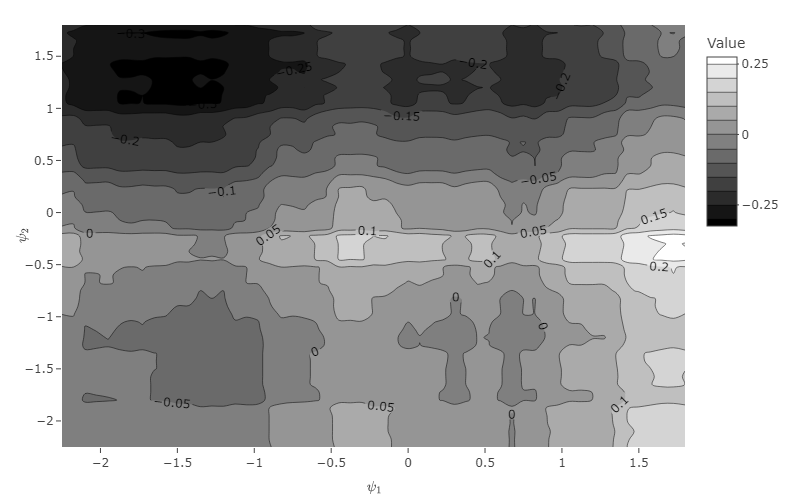}
		\subcaption{}
	\end{subfigure}
	\caption{Simulation II: (a) Value function  (b) IPW-surface.} 
	\label{simulII_value}
\end{figure}

Simulation II explores a two-stage treatment rule, treat if $x_k>\psi_k $, $ \psi_k \in [-2.25,1.8]$, $ k=1,2 $. This example examines a value function that is multi-modal, with one global maximizer, and some other local maxima. The data-generating mechanism for this simulation is as follows:
\begin{align*}
	y=0.2x_1-&0.2(x_1+2.25)(x_1+1.5)(x_1+0.3)(x_1-1.8)(x_1-.75)(\mathbbm{1}_{(x_1-1.5)>0}-z_1)\\
			-&0.2(x_2+2.1)(x_2+1.65)(x_2+0.3)(x_2-2.1)(x_2-1.35)(\mathbbm{1}_{(x_2-0.75)>0}-z_2)+\epsilon.
\end{align*}
Intermediary variables are distributed as $x_1 \sim N(0,1.5^2) $; $ x_2\sim1.5z_1+N(0,1.5^2)$ and treatment variables as $ z_1\sim Bern(expit(-(1/1.5)x_1)) $ and $ z_2\sim Bern(expit(-(1/1.5)x_2+(1/1.5)z_1) $. Additive noise is distributed as $ \epsilon \sim N(0,0.3^2) $. In Appendix D, we also explore heteroskedastic additive noise. 
The value function is given in Figure \ref{simulII_value} (a), with 3-D version found in the  \href{https://danroduq.github.io/ObjectiveII/index.html#value-function}{Interactive Supplement}. As in Simulation I, this problem exhibits multi-modality, thus we focus on medians and IQRs. From Figure \ref{simulII_value} (b) we observe the IPW-surface still captures the general characteristics of the value function. This can also be seen in the \href{https://danroduq.github.io/ObjectiveII/index.html#ipw-surface-n500}{Interactive Supplement}. 

\begin{figure}[H]
	\centering
	\hspace{-1.3cm}
	\begin{subfigure}[b]{0.3\linewidth} 
		\includegraphics[scale=0.23]{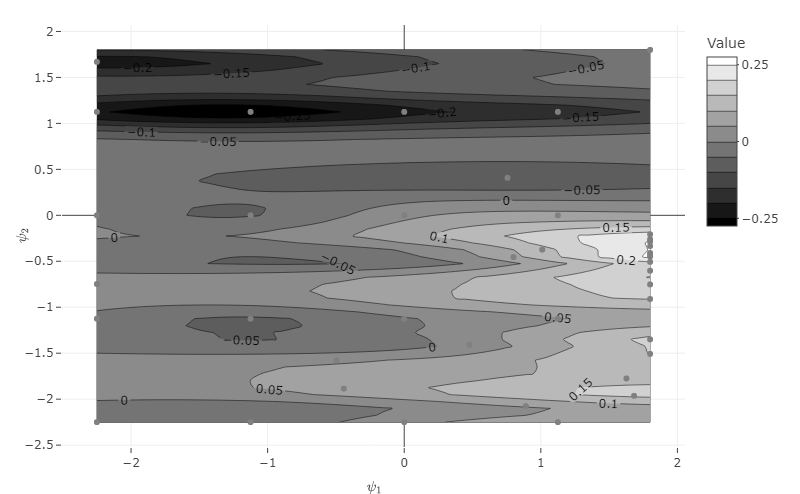}
		\caption{}
	\end{subfigure}
	\begin{subfigure}[b]{0.3\linewidth} 
		\includegraphics[scale=0.23]{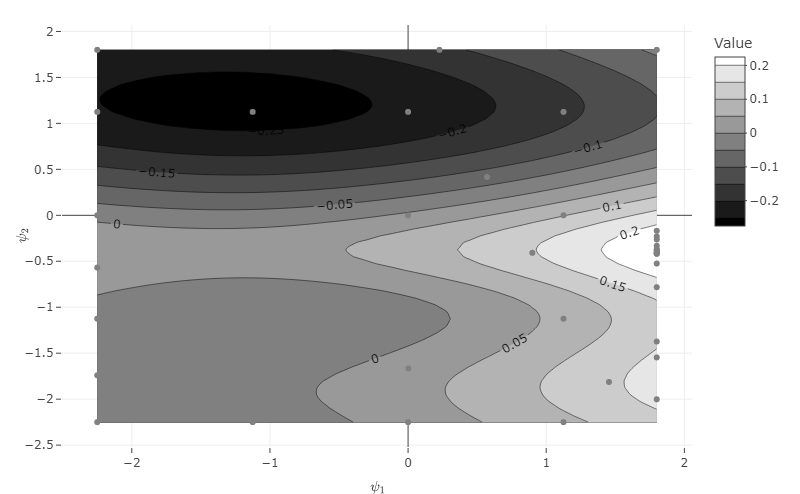}
				\caption{}
	\end{subfigure}
	\begin{subfigure}[b]{0.3\linewidth} 
		\includegraphics[scale=0.23]{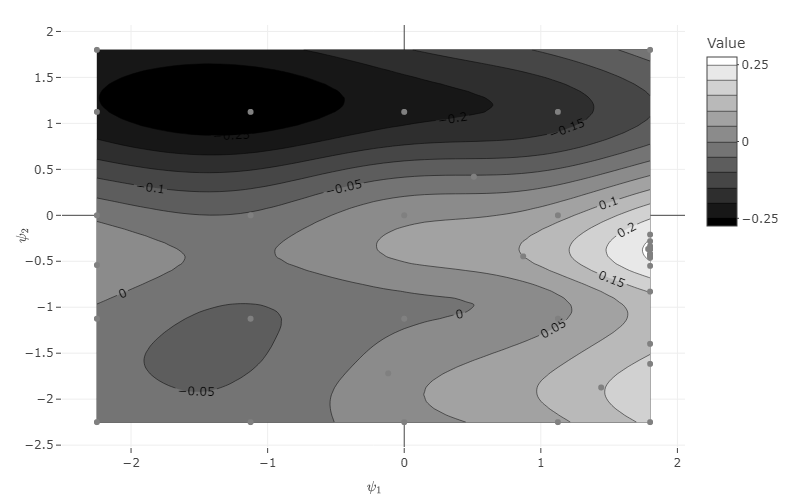}
				\caption{}
	\end{subfigure}
	\caption{Simulation II: Contour plot of emulation surface at +25 points (a) Int$ \mathcal{GP} $   (b) HM$ \mathcal{GP} $  (c) HE$ \mathcal{GP} $.} 
	\label{25_n500_cov3_prior0_homosk}
\end{figure}

An initial set of design points is taken in increments of 0.75 to yield at a total of 16 points. Before examining the results for each of these settings, we examine the results of a grid-search. From Table \ref{simulII_grid}, we see that there is a high amount of variability in the estimated optimal $ \psi_1 $ parameter, as measured by the IQR. This is similar to what was observed in Simulation I. Estimates of $ \psi_2 $ perform better, as there is no multi-modality in this axis. In what follows, we will compare the $ \mathcal{GP} $ approaches to the grid-search. 
\begin{table}[H]
	\centering
	\caption{ Simulation II: Results for grid-search with increments of 0.05 and $ n=500 $. True $ (\psi_{1opt},\psi_{2opt})=(1.8,-0.3) $; true value at optimum 0.241.} 
	\begin{tabular}{rlll}
		\hline
		Statistic & $ \hat{\psi}_{1opt} $ & $ \hat{\psi}_{2opt} $ & Value at Optimum \\ 
		\hline
		Mean (SD)   &   1.098  (1.140) & -0.409  (0.382)  & 0.277  (0.094)\\ 
		Median (IQR) &  1.725 (1.725)  &-0.375  (0.300)  &0.275  (0.132)\\ 
		\hline
	\end{tabular}
	\label{simulII_grid}
\end{table}
 From Figure \ref{25_n500_cov3_prior0_homosk}, which shows one replicate analysis for each of the three $ \mathcal{GP} $ methods, we see from the points on the plot that the cross-section at $ \psi_1=1.8 $ is explored the most; this cross-section contains the global optimizer. For this replicate, the second optimum is not well identified by any of the $ \mathcal{GP}s $. 

\begin{figure}[H]
	\centering
	\hspace{-1.3cm}
	\begin{subfigure}[b]{0.5\linewidth} 
		\includegraphics[scale=0.25]{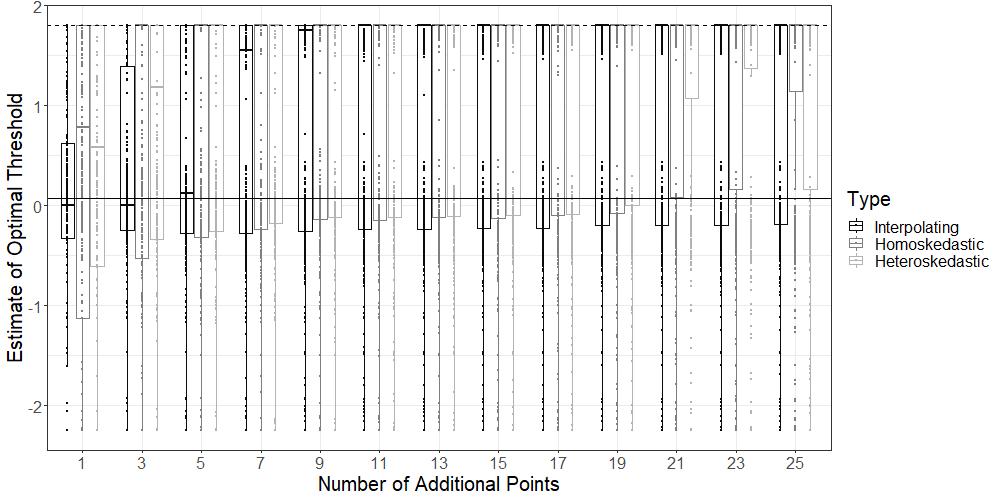}
		\caption{}
	\end{subfigure}
	\begin{subfigure}[b]{0.5\linewidth} 
		\includegraphics[scale=0.25]{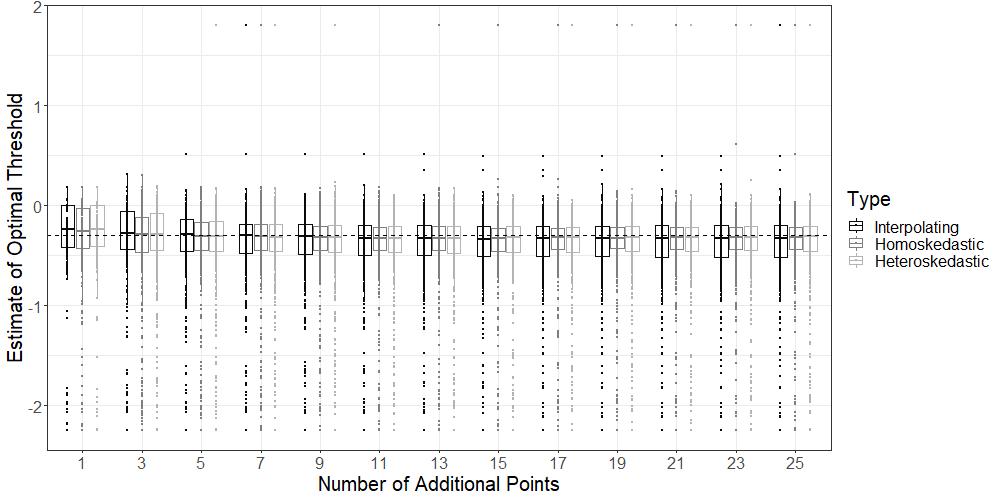}
		\caption{}
	\end{subfigure}
	\caption{Simulation II: Boxplot at $ +m $ points; $ n=500 $ with 16 design points: (a) Optimal $ \psi_1 $  (b) Optimal $ \psi_2 $.} 
	\caption{}
	\label{simulII_boxplotpsis_n500_cov3_prior0}
\end{figure}

\begin{figure}[H]
	\centering
	\includegraphics[scale=0.25]{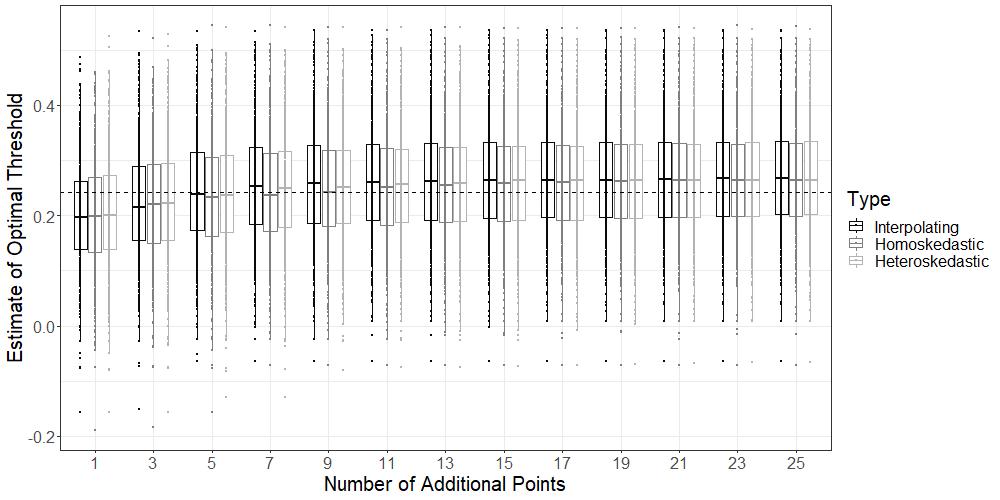}
		\caption{Simulation II: Boxplot of value at optimum after $ +m $; n=500 with 16 design points.} 
	\label{simulII_boxplotvalue_n500_cov3_prior0}
\end{figure}

Comparing Table \ref{simulII_tab_medianpsi_n500_cov3_prior0} with the results of the grid-search, we note that at $ +25$ points the median optimal values resulting from the HM$\mathcal{GP}$ are closer to the truth than those arrived at via a grid-search; most notably the IQR  for $ \psi_{1opt} $ is much smaller. This strengthens the observation from Simulation I that a grid-search is not always the most robust approach. We also observe that the HE$\mathcal{GP}$ outperforms the grid-search at $ +25 $ points. In the $ \psi_2 $ direction, all three methods perform similarly, with the HM$\mathcal{GP}$ and HE$\mathcal{GP}$ outperforming the grid-search at 25 additional samples. We see from Table \ref{simulII_tab_medianvalue_n500_cov3_prior0} that all $ \mathcal{GP} $s perform equally well in estimating the value at the optimum. From Figure \ref{simulII_boxplotpsis_n500_cov3_prior0}  we can visualize how sampling additional experimental points improves the estimation of $ \psi_{1opt} $ and $ \psi_{2opt} $. From panel (a), we see that after 11 sampled points, the first quartile and the median $ \hat{\psi}_{1opt} $ are at the true value of the optimal threshold for all $ \mathcal{GP} $ methods. The solid horizontal line on the plot is placed at the grid-search $ IQR+\psi_{1opt} $ value. This allows us to see that after $21$ sampled points the IQR for both the HM$ \mathcal{GP} $ and the HE$ \mathcal{GP} $ is smaller or equal to that of the grid-search IQR which uses $ 3721 $ grid points, thereby emphasizing the increased efficiency that a $ \mathcal{GP} $ approach can provide. We note additionally that the HE$\mathcal{GP}$ achieves improved results slightly faster than the HM$ \mathcal{GP} $, however the HM$\mathcal{GP}$ achieves comparable results after a few additionally sampled points. Panel (b) in the plot shows the results for the $ \psi_{2opt} $ parameter; we see that all $ \mathcal{GP} $ methods perform consistently well, with the HM$ \mathcal{GP} $ and HE$ \mathcal{GP} $ performing slightly better than the Int$ \mathcal{GP} $. From Figure \ref{simulII_boxplotvalue_n500_cov3_prior0}, we see that the estimation of the optimal value is similar across all methods. 
Overall, this example allows us to conclude that a method that acknowledges noise in the estimation surface is important in order to more precisely estimate the optimizers. We note again that the improvement offered by the $ \mathcal{GP} $ is most relevant in the direction of multi-modality. In Appendix D, we see that an increased sample size improves the estimation of the $ \psi_{1opt} $ parameter, but that the HM$ \mathcal{GP} $ and the HE$ \mathcal{GP} $ still outperform the grid-search and the interpolating approach. Although the HM$ \mathcal{GP} $ seemed to require slightly more data to achieve the performance of the HE$\mathcal{GP} $ for this specific setting, we must keep in mind that the HE$\mathcal{GP} $ is more computationally intensive than the HM$\mathcal{GP} $ approach as it requires fitting a $ \mathcal{GP} $ on estimated residuals. We note additionally that in all simulations shown in Appendix D, the performance of the HM$ \mathcal{GP} $ and HE$\mathcal{GP} $ is comparable

\begin{table}[H]
	\centering
	\caption{Simulation II: Optimal $ \psi_1 $ and $ \psi_2 $ after $ +m $ points, median (IQR); $ n=500 $ with 16 design points. True $  (\psi_{1opt},\psi_{2opt})=(1.8,-0.3)  $.} 
	\begin{tabular}{clllllll}
		\hline
	Parameter & Method & +1 & +5 & +10 & +15 & +20 & +25 \\ 
		\hline
		$ \psi_{1opt} $ & Int$ \mathcal{GP} $ & 0.000 (0.950) & 0.123 (2.074) & 1.800 (2.036) & 1.800 (2.026) & 1.800 (1.995) & 1.800 (1.988) \\ 
		$ \psi_{1opt} $ & HM$ \mathcal{GP} $  & 0.779 (2.925) & 1.800 (2.116) & 1.800 (1.930) & 1.800 (1.928) & 1.800 (1.805) & 1.800 (0.657) \\ 
		$ \psi_{1opt} $ & HE$ \mathcal{GP} $  & 0.580 (2.406) & 1.800 (2.054) & 1.800 (1.925) & 1.800 (1.901) & 1.800 (1.731) & 1.800 (1.636) \\ 
		\hline
		$ \psi_{2opt} $ &Int$ \mathcal{GP} $ & -0.241 (0.420) & -0.285 (0.325) & -0.321 (0.296) & -0.334 (0.301) & -0.317 (0.311) & -0.331 (0.318) \\ 
		$ \psi_{2opt} $ &HM$ \mathcal{GP} $  & -0.256 (0.400) & -0.306 (0.286) & -0.322 (0.242) & -0.328 (0.225) & -0.317 (0.216) & -0.318 (0.219) \\ 
		$ \psi_{2opt} $ &HE$ \mathcal{GP} $  & -0.242 (0.411) & -0.312 (0.300) & -0.327 (0.260) & -0.319 (0.262) & -0.323 (0.249) & -0.312 (0.247)\\
		\hline
	\end{tabular}
		\label{simulII_tab_medianpsi_n500_cov3_prior0}
\end{table}

\begin{table}[H]
		\centering
		\caption{Simulation II: Value at optimum after $ +m $ points, median (IQR); n=500 with 16 design points. True value at optimum: 0.241.} 
	\begin{tabular}{rllllll}
		\hline
		& +1 & +5 & +10 & +15 & +20 & +25 \\ 
		\hline
		Int$ \mathcal{GP} $ & 0.196 (0.123) & 0.238 (0.141) & 0.258 (0.140) & 0.264 (0.138) & 0.264 (0.136) & 0.267 (0.133) \\ 
		HM$ \mathcal{GP} $  & 0.198 (0.136) & 0.234 (0.144) & 0.247 (0.141) & 0.259 (0.136) & 0.264 (0.137) & 0.265 (0.131) \\ 
		HE$ \mathcal{GP} $  & 0.200 (0.135) & 0.236 (0.140) & 0.255 (0.133) & 0.263 (0.135) & 0.264 (0.133) & 0.264 (0.132) \\ 
		\hline
	\end{tabular}
	\label{simulII_tab_medianvalue_n500_cov3_prior0}
\end{table}

\subsection{Simulation III}

For simulation III, we explore a family of regimes indexed by $\psi_1,\psi_2,\psi_3$ such that $\psi_{1}x_{k1}+\psi_{2}x_{k2}>0.5-3\psi_3u; \;  k=1,...,4;$  $ x_{k1}, x_{k2} $ are Normally distributed intermediate covariates and $ u $ is a binary baseline covariate. Details of the data-generating mechanism can be found in Appendix E. In the results tables, we do not include a $ \psi_2 $ column, as we apply the following constraint: $ \psi_1+\psi_2=1 $. Note that $ \psi_1,\psi_2\in[0.2,0.8]$ and $\psi_3 \in[-0.3,0.3] $. The known optimizer is $(\psi_1,\psi_3)=(0.5,0.1) $ and the value at the optimizer is $ 1 $. A set of 20 design points is obtained by sampling in increments of $ 0.2 $ and $ 0.15 $ in $ \psi_1 $ and $ \psi_3 $ directions.

\begin{figure}[H]
	\centering
	\begin{subfigure}[b]{0.45\linewidth} 
		\includegraphics[scale=0.25]{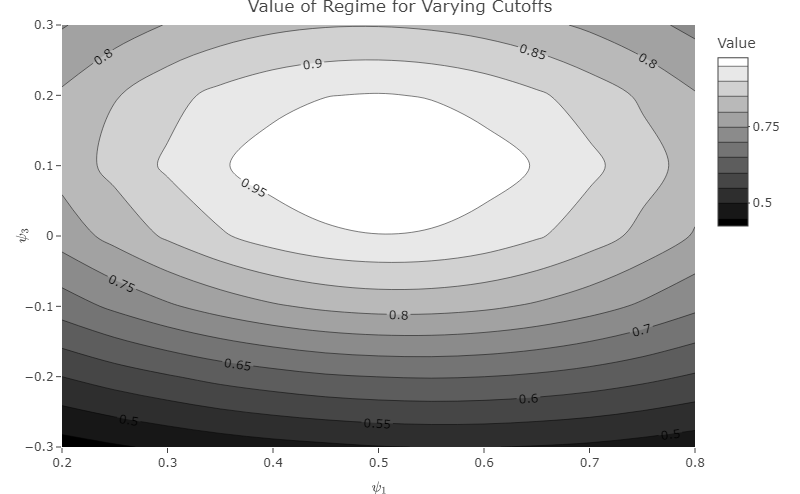}
				\caption{}
	\end{subfigure}
	\begin{subfigure}[b]{0.45\linewidth} 
		\includegraphics[scale=0.25]{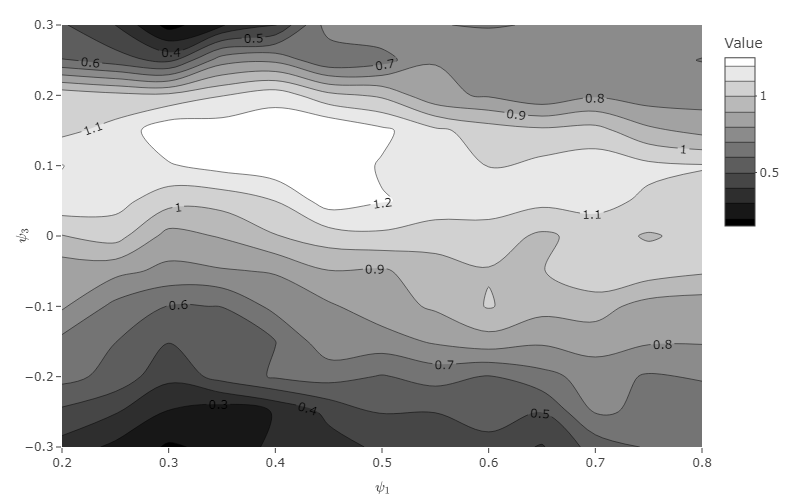}
				\caption{}
	\end{subfigure}
	\caption{Simulation III: (a) Value function  (b) Estimated value function using normalized IPW.} 
	\label{simulIII_value}
\end{figure}

\begin{figure}[H]
	\centering
	\hspace{-1.3cm}
	\begin{subfigure}[b]{0.3\linewidth} 
		\includegraphics[scale=0.23]{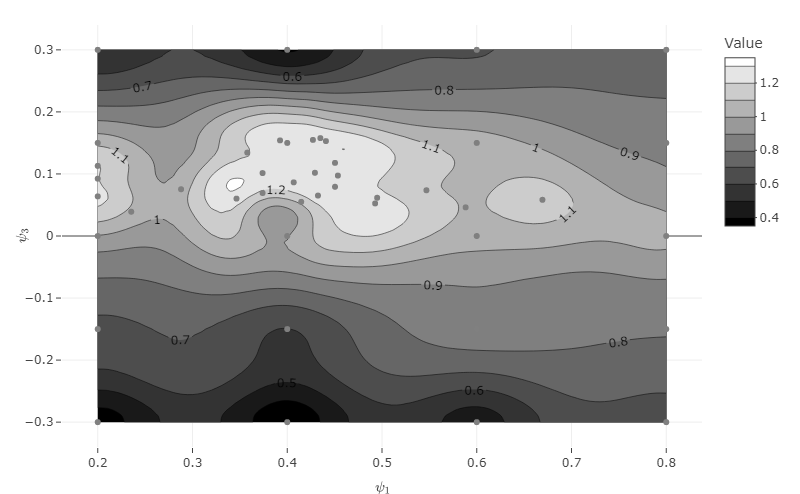}
				\caption{}
	\end{subfigure}
	\begin{subfigure}[b]{0.3\linewidth} 
		\includegraphics[scale=0.23]{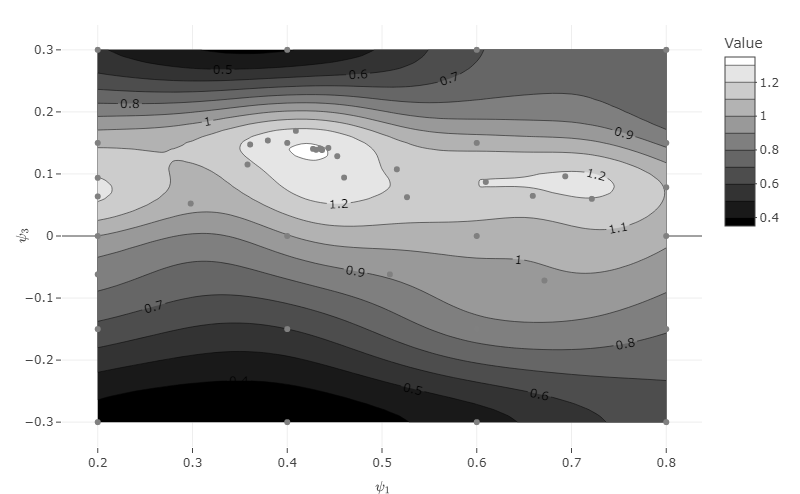}
				\caption{}
	\end{subfigure}
	\begin{subfigure}[b]{0.3\linewidth} 
		\includegraphics[scale=0.23]{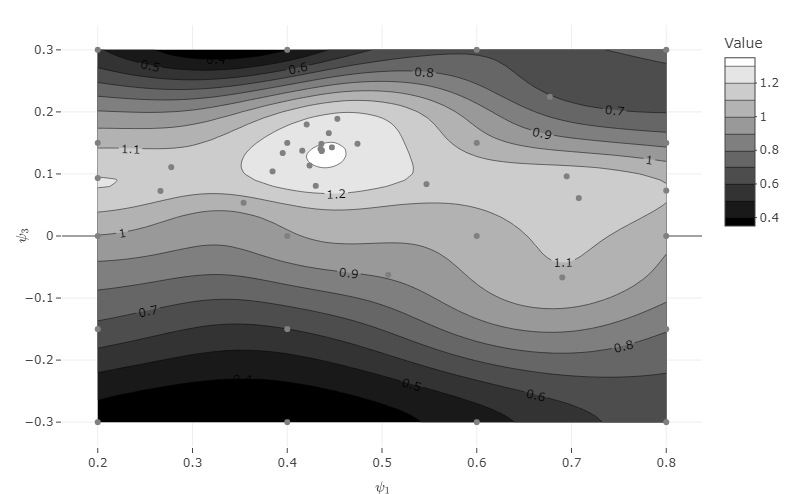}
				\caption{}
	\end{subfigure}
	\caption{Simulation III: Contour plot of emulation surface at +25 points: (a) Int$ \mathcal{GP} $   (b) HM$ \mathcal{GP} $ (c) HE$ \mathcal{GP} $.} 
	\label{simulIII_25_n500_prior0_homosk}
\end{figure}

We see from Figure \ref{simulIII_value} (a) that this is a uni-modal setting, different from Simulation I and II. From Figure \ref{simulIII_value} (b) we see how the IPW-surface captures the general form of the value function. Although this is a uni-modal example, in what follows, our presentation focuses on medians and interquartile ranges, in order to maintain consistency with the other simulations. Additional tables can be found in Appendix E.
From Table \ref{simulIII_grid}, we see that the grid-search performs better than in the multi-modal examples, with variability around the optimizer decreasing. We observe from Figure \ref{simulIII_25_n500_prior0_homosk} that for a fixed replicate, the HE$ \mathcal{GP} $ best captures the shape of the true value function. 
\begin{table}[H]
	\centering
	\caption{Simulation III: Grid-search results in increments of $ 0.01 $ and $ n=500 $. True $ (\psi_{1opt}, \psi_{3opt})=(0.5,0.1) $; true value at optimum: 1.} 
	\begin{tabular}{rlll}
		\hline
		Statistic & $ \hat{\psi}_{1opt} $ & $ \hat{\psi}_{3opt} $ & Value at Optimum\\ 
		\hline
		  Mean (SD)   & 0.471 (0.153) & 0.104 (0.120) & 1.233 (0.147) \\
		 Median (IQR) & 0.470 (0.220) & 0.110 (0.150) & 1.231 (0.189) \\ 
		\hline
	\end{tabular}
	\label{simulIII_grid}
\end{table}

\begin{figure}[H]
	\centering
	\hspace{-1.3cm}
	\begin{subfigure}[b]{0.5\linewidth} 
		\includegraphics[scale=0.3]{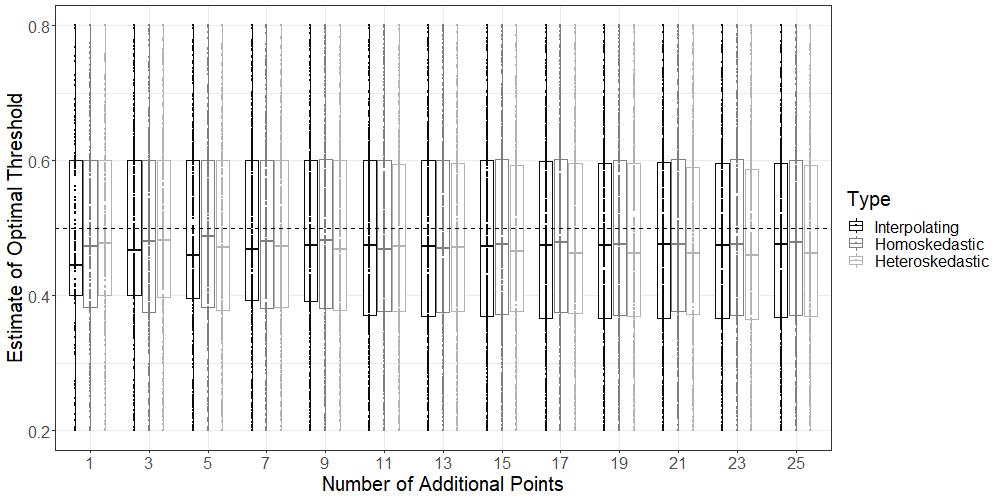}
				\caption{}
	\end{subfigure}
	\begin{subfigure}[b]{0.5\linewidth} 
		\includegraphics[scale=0.3]{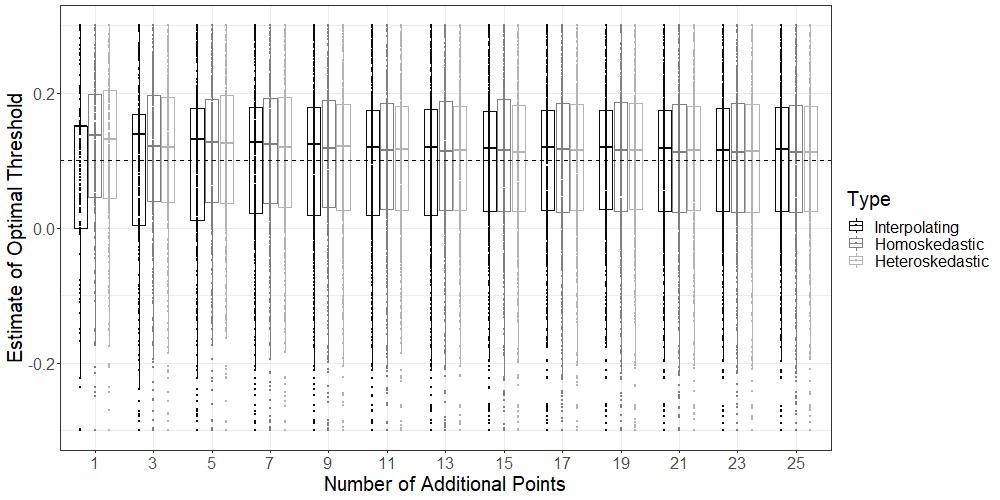}
				\caption{}
	\end{subfigure}
	\caption{Simulation III: Boxplot after $ +m $ points; $ n=500 $ with 20 design points (a) Optimal $ \psi_1 $  (b) Optimal $ \psi_3 $.} 
	\label{simulIII_psi_box_n500}
\end{figure}

\begin{figure}[H]
	\centering
	\includegraphics[scale=0.3]{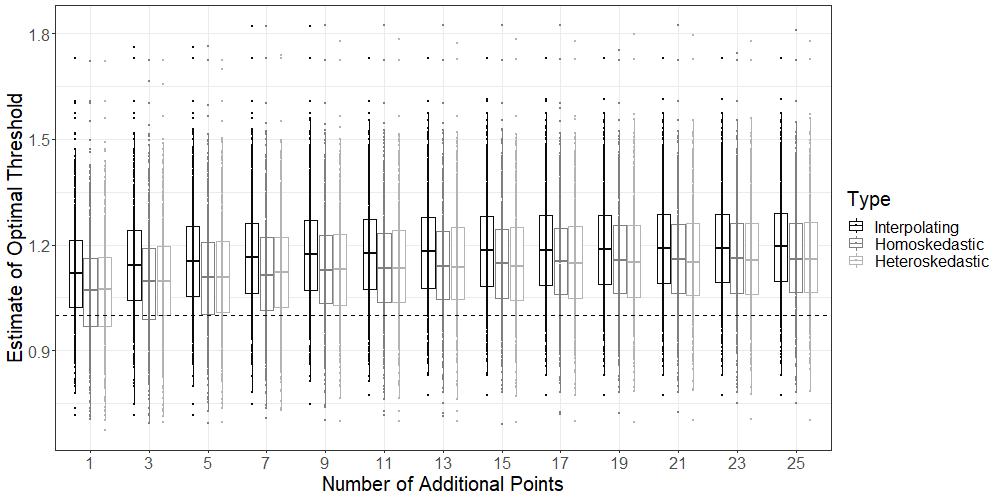}
	\caption{Simulation III: Boxplot of value at optimum after $+m$; n=500 with 20 design points.} 
	\label{simulIII_value_box_cov3_n500}
	\end{figure}

  From Table \ref{simulIII_tab_medianpsis_cov3_n500_prior0}, we see that all three $ \mathcal{GP} $s yield good results, even for a small number of additional samples. Given the results of simulations I and II, these results suggest that the choice of $ \mathcal{GP}  $ modeling approach is most consequential in multi-modal settings and that there is no drawback in using a HM$ \mathcal{GP} $, even if the value function is uni-modal. In a real-data analysis we do not have knowledge of whether we are in a multi-modal problem; hence, a $ \mathcal{GP} $ approach that acknowledges variability in the estimation surface is advisable. We note as well that all  three $\mathcal{GP}$ modeling approaches achieve good performance for a small fraction of the function evaluations required by a grid-search. Figure \ref{simulIII_psi_box_n500} shows that even at a few additional points, the optimizers are well identified. From Table \ref{simulIII_tab_medianvalues_cov3_n500_prior0}, we see that as additional points are sampled, the accuracy of the estimated optimal value decreases slightly; this can also be observed in Figure \ref{simulIII_value_box_cov3_n500}. Three dimensional renderings related to this simulation can be found in \href{https://danroduq.github.io/ObjectiveII/simulation-iii-main-paper.html}{Interactive Supplement}.
  
\begin{table}[H]
	\centering
	\caption{Simulation III: Optimal $ \psi_1 $ and $ \psi_3 $ after $ +m $ points, median (IQR); $ n=500 $ with 20 design points. True $(\psi_{1opt},\psi_{3opt})=(0.5, 0.1) $.} 
	\begin{tabular}{clllllll}
		\hline
		Parameter& Method & +1 & +5 & +10 & +15 & +20 & +25 \\ 
		\hline
		$ \psi_{1opt} $ & Int$ \mathcal{GP} $  & 0.445 (0.200) & 0.460 (0.204) & 0.476 (0.216) & 0.473 (0.231) & 0.476 (0.230) & 0.476 (0.229) \\ 
		$ \psi_{1opt} $ & HM$ \mathcal{GP} $   & 0.473 (0.217) & 0.488 (0.218) & 0.471 (0.224) & 0.475 (0.230) & 0.475 (0.228) & 0.479 (0.230) \\ 
		$ \psi_{1opt} $ & HE$ \mathcal{GP} $   & 0.477 (0.200) & 0.471 (0.223) & 0.467 (0.219) & 0.466 (0.217) & 0.462 (0.226) & 0.462 (0.224) \\ 
		\hline
		$ \psi_{3opt} $ & Int$ \mathcal{GP} $ & 0.150 (0.150) & 0.131 (0.166) & 0.121 (0.159) & 0.118 (0.148) & 0.118 (0.150) & 0.116 (0.155) \\ 
		$ \psi_{3opt} $ & HM$ \mathcal{GP} $  & 0.137 (0.152) & 0.127 (0.153) & 0.117 (0.164) & 0.115 (0.167) & 0.115 (0.159) & 0.112 (0.159) \\ 
		$ \psi_{3opt} $ & HE$ \mathcal{GP} $  & 0.131 (0.159) & 0.125 (0.160) & 0.119 (0.156) & 0.113 (0.158) & 0.116 (0.158) & 0.112 (0.155) \\ 
		\hline
	\end{tabular}
	\label{simulIII_tab_medianpsis_cov3_n500_prior0}
\end{table}

\begin{table}[H]
	\centering
	\caption{Simulation III: Value at optimum after $ +m $ points, median (IQR); $ n=500 $ with 20 design points. True value at optimum: 1.} 
	\begin{tabular}{rllllll}
		\hline
		& +1 & +5 & +10 & +15 & +20 & +25 \\ 
		\hline
		Int$ \mathcal{GP} $ & 1.118 (0.189) & 1.154 (0.200) & 1.174 (0.198) & 1.185 (0.200) & 1.187 (0.197) & 1.195 (0.193) \\ 
		HM$ \mathcal{GP} $  & 1.070 (0.194) & 1.108 (0.205) & 1.128 (0.202) & 1.147 (0.196) & 1.156 (0.197) & 1.160 (0.196) \\ 
		HE$ \mathcal{GP} $  & 1.074 (0.196) & 1.108 (0.200) & 1.129 (0.205) & 1.138 (0.208) & 1.148 (0.204) & 1.158 (0.201) \\ 
		\hline
	\end{tabular}
	\label{simulIII_tab_medianvalues_cov3_n500_prior0}
\end{table}

\subsection{Case Study}

In this case study, we analyze data from \cite{hammer1996trial} to illustrate how the $ \mathcal{GP} $ methods may be applied in a setting with real data. These data come from a double-blinded randomized trial of HIV antiretroviral therapies undertaken to compare treatments using single and double nucleosides as a means of treating HIV type 1. Patients with CD4 cell counts between 200 to 500 $ cells/\mu L $  were enrolled in the study. A total of 2467 patients were assigned to one of four daily regimens 1) 600 mg of zidovudine, 2) 600 mg of zidovudine \& 400 mg of didanosine, 3) 600mg of zidovudine \& 2.5 mg zalcitabine, or 4) 400 mg didanosine. The primary end-point in the study was an observation of $ \ge 50 $ percent decline in CD4 cell count, progression to AIDS, or death. Overall, the zidovudine regimen was found to be inferior to other regimens, with regard to the primary end-point. 

 Variables found in the dataset include those captured at baseline including patients' race, sex, baseline CD4, weight, age, history of antiretroviral therapy, symptoms of HIV infection, and  Karnofsky score, as well as those captured later in the study such as 20 week CD4. These data may be accessed via the \textit{LongCART} package in \textit{R} \citep{kundu2021package}. We restrict our analysis to the use of two dual-therapies and aim to determine which patients should be given zidovudine with zalcitabine versus zidovudine with didanosine, thereby recognizing that mono-therapy is widely considered inadequate. In particular, we examine whether tailoring on baseline CD4 cell count and baseline weight yields improved 20 week CD4 cell counts, the outcome of interest. There are no missing data in any of the variables required for this analysis. There are 524 patients in the zidovudine \& zalcitabine arm and 522 in the zidovudine \& didanozine arm. The known treatment probability is $ 0.5 $ by design, however we estimate these probabilities, as this has been shown to improve efficiency when using IPW estimators \citep{henmi2004paradox}. Now, the specific family of regimes that we consider is: receive zidovudine with didanosine if  $ \text{baseline weight}> \psi_W$ and  $ \text{baseline CD4}>\psi_{CD4}$, where $ \psi_W\in [50,100] $ and $\psi_{CD4} \in [200,600]$.  For every regime index $ (\psi_W, \psi_{CD4}) $, a value is estimated, and this is used to inform the resulting $ \mathcal{GP} $, regardless of whether it is a one-stage decision rule or a multi-stage decision rule. In this analysis, we make use of the normalized IPW estimator for the value of a regime, and pair it with the proposed $ \mathcal{GP} $ approaches.

\begin{figure}[H]
	\centering
	\begin{subfigure}[b]{0.4\linewidth} 
		\includegraphics[scale=0.3]{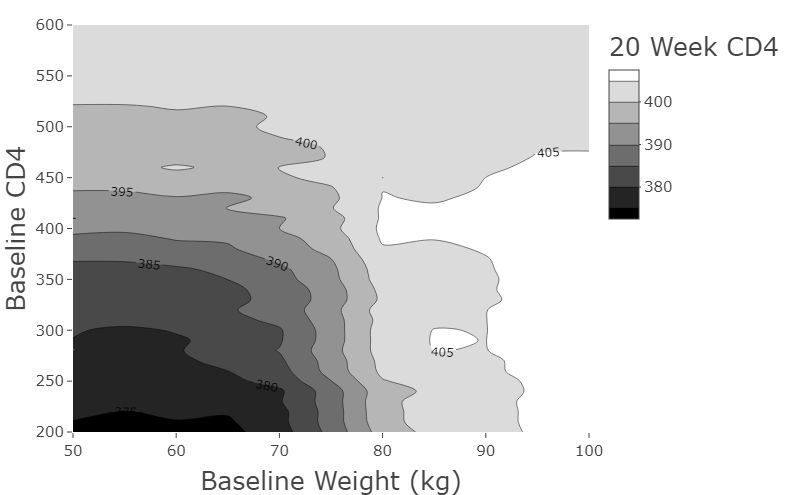}
	\end{subfigure}\\
	\caption{ Contour plot of normalized IPW-surface.}
	\label{ValueFunci}
\end{figure}

Using the normalized IPW estimator on a fine grid of points yields the value function displayed in Figure \ref{ValueFunci}. We see that there appears to be a trough for combinations of low $ \psi_W $ and low $  \psi_{CD4} $; there is also a high value region for large $ \psi_W $, across a wide range of $ \psi_{CD4} $. From the 3-D rendering in the \href{https://danroduq.github.io/ObjectiveII/case-study.html#normalized-ipw-surface}{Interactive Supplement}, we see that the IPW-surface is rather smooth; this, in part, is brought about by the use of the normalized IPW. We have also examined the resulting surface when using the standard IPW estimator, and it was characteristically more noisy, leading to the possibility of more modeling challenges. 
A standard approach that one may take via value-search estimation is to perform a grid-search for the optimal regime $(\psi^W_{opt},  \psi^{CD4}_{opt})$. It is noteworthy that for a single sample, as in this analysis, a grid-search for the optimal regime will not provide a measure of sampling uncertainty around the identified optimum. To arrive at a complete statistical analysis of these data, we seek to quantify this uncertainty by using the Bayesian bootstrap over 500 samples, where at each sample a Dirichlet vector is observed, with all concentration parameters equal to one and dimension equal to the number of patients in the study. If, for each of these bootstrap samples, we compute the optimal regime, then what results is a distribution for the optimum. We report the median optimal index and optimal value, with $ 95\% $ credible intervals. We do this for both a coarse and fine grid to examine whether the grid choice impacts the results, which are shown in Table \ref{Tab:GridSearch}. The coarse grid has increments of $ 15 \; kg $ and $ 35\; cells/\mu L  $ in the weight and CD4 axes, respectively; the fine grid has increments of $ 10 \; kg $ and $ 20 \; cells/\mu L $. We see from the table that as expected, the choice of grid impacts the resulting inference. Table \ref{Tab:GridSearch} also shows the results of fitting a quadratic MSM with mean ($ \beta_0 +\beta_1\psi_W+\beta_2\psi_W^2+ \beta_3\psi_{CD4}+\beta_4\psi_{CD4}^2 +\beta_5\psi_W\psi_{CD4}$), using the same bootstrapping approach. The fitted model appears to fit the data relatively well, as shown in \href{https://danroduq.github.io/ObjectiveII/case-study.html#value-surface-using-msm}{Interactive Supplement}. However, note that there is no variability in the optimal $\psi_W $, revealing some deficiencies in the model.  

\begin{figure}[H]
	\centering
	\hspace{-1.3cm}
	\begin{subfigure}[b]{0.3\linewidth} 
		\includegraphics[scale=0.14]{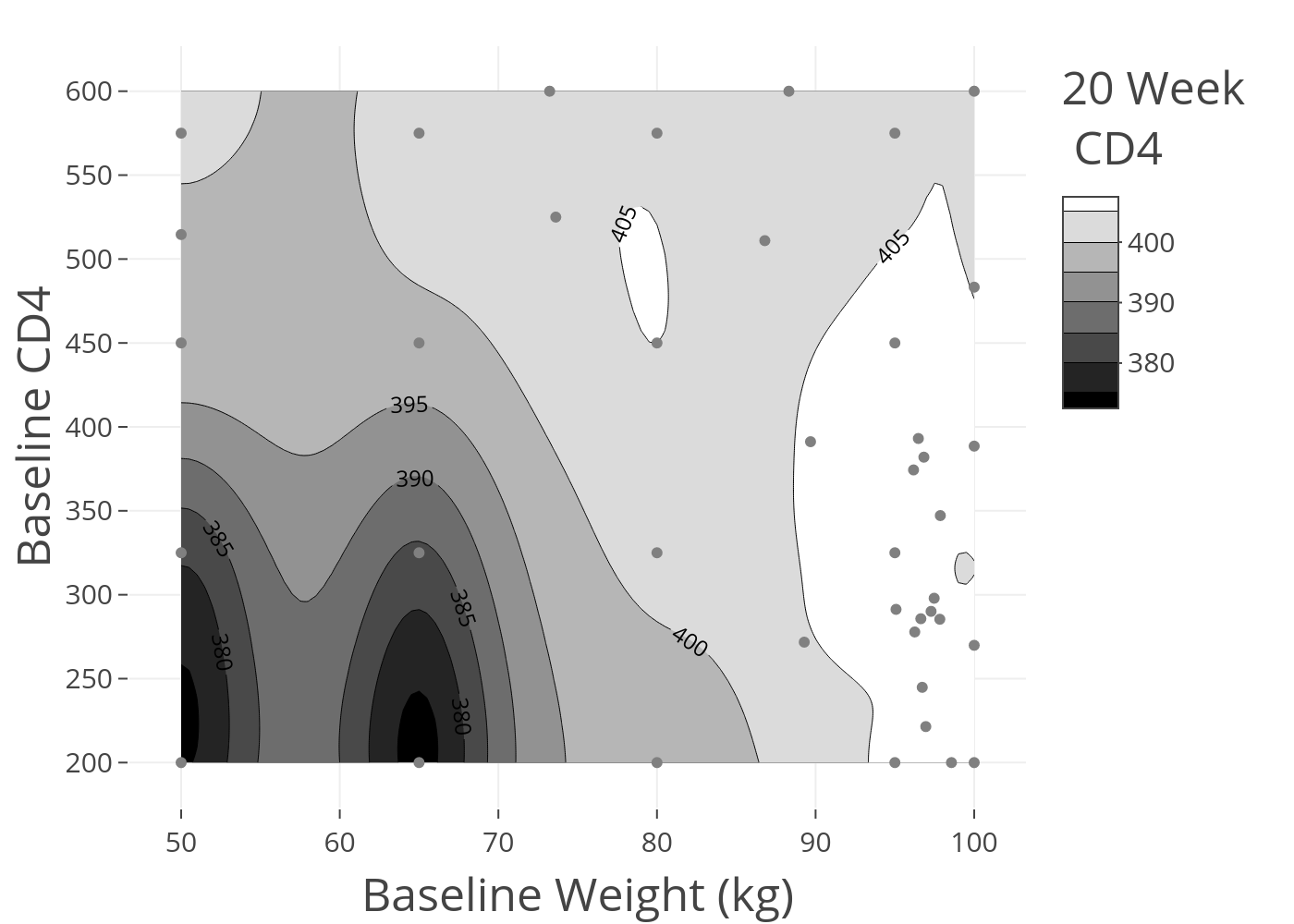}
		\caption{}
	\end{subfigure}
	\begin{subfigure}[b]{0.3\linewidth} 
		\includegraphics[scale=0.14]{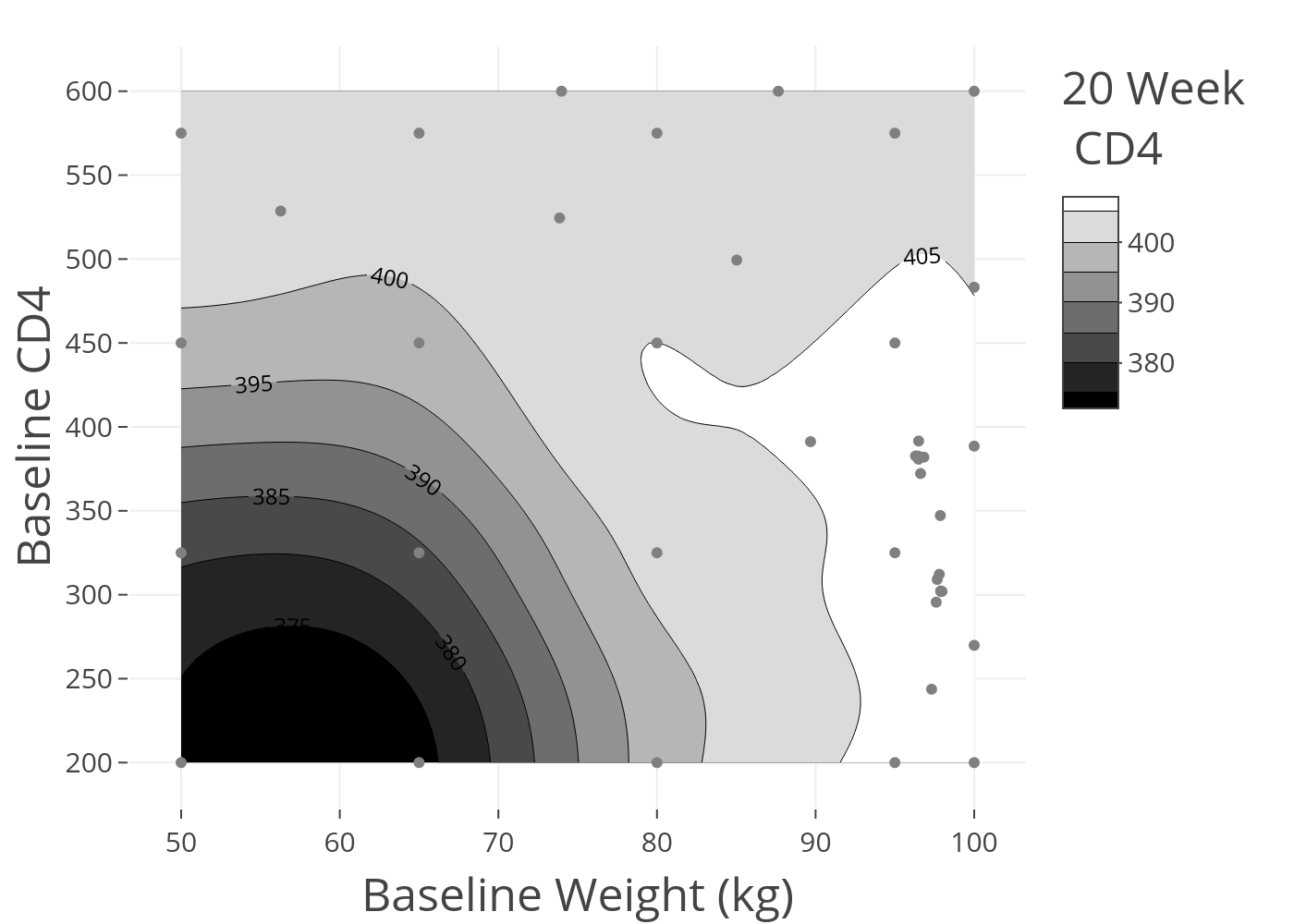}
		\caption{}
	\end{subfigure}
	\begin{subfigure}[b]{0.3\linewidth} 
		\includegraphics[scale=0.14]{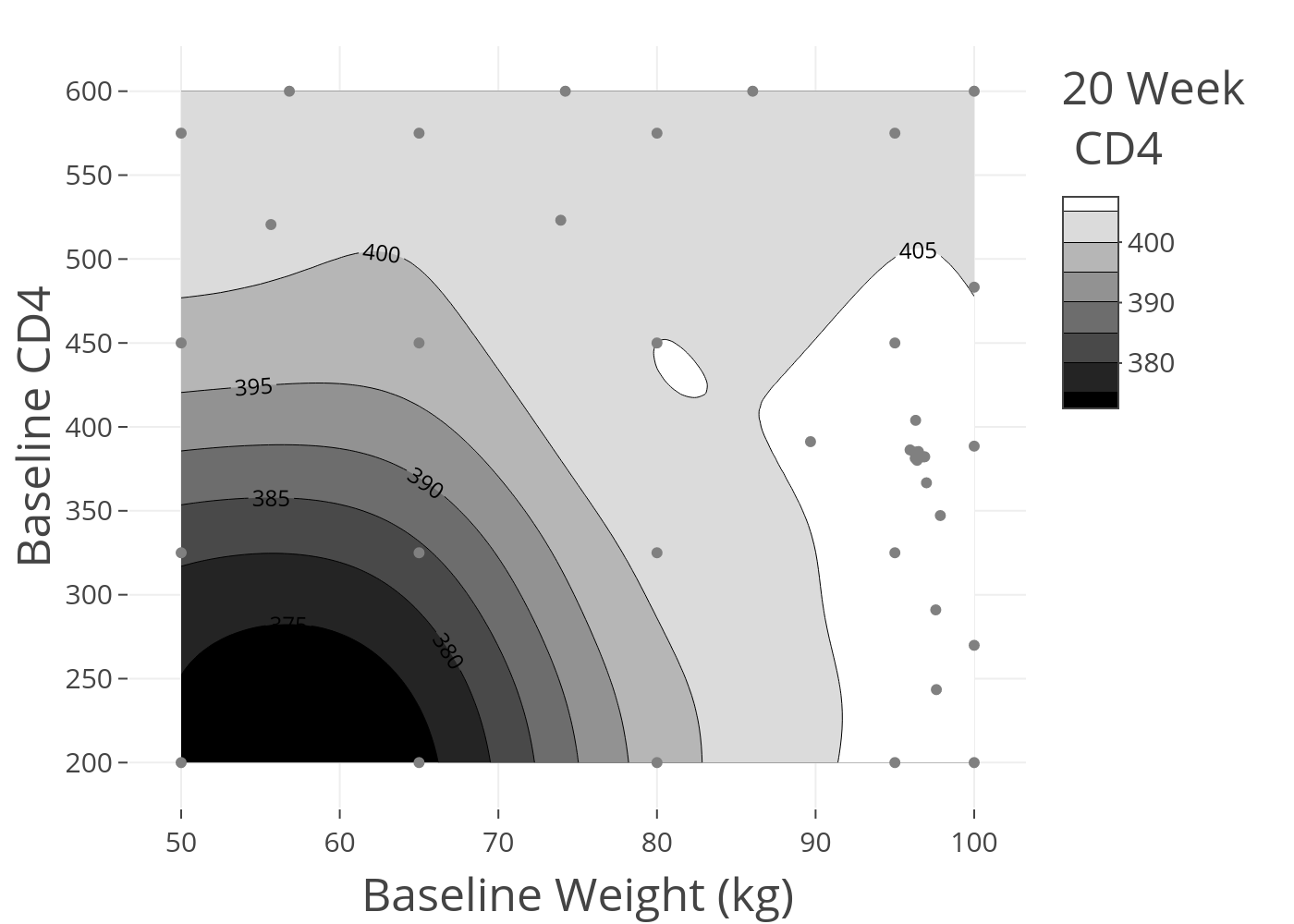}
		\caption{}
	\end{subfigure}
	\caption{HIV Study: Contour plot of emulation surface at +25 points (a) Int$ \mathcal{GP} $ (b) HM$ \mathcal{GP} $  (c) HE$ \mathcal{GP} $.} 
	\label{CaseStudyResulties}
\end{figure}

In addition to sampling uncertainty, there is another type of uncertainty that can be important to quantify in the grid-search approach. This is uncertainty reflecting the coarseness of the grid chosen. For a fixed grid, the identified optimum has uncertainty relative to the optimum that would be found were we to use a grid with increments approaching zero. However, there is no clear way to incorporate this uncertainty using a grid-search. As we will discuss, the $ \mathcal{GP} $ approach can attribute more uncertainty to regions that have not been well explored and combine this with the sampling uncertainty.
\begin{table}[H]
	\centering
	\caption{HIV Study: Estimated optimal value and optimal index via a coarse and fine grid-search, with 95\% credible intervals.}
	\begin{tabular}{rlll}
		\hline
		 Type & Coarse Grid & Fine Grid & MSM\\ 
		\hline
		 $ \hat{\psi}^{CD4}_{opt} $  & 305 (200-533)   & 280 (200-460) & 343 (200-440)\\ 
		 $ \hat{\psi}^{W}_{opt} $    &  95 (80-95)     & 100 (80-100)  & 100 (100-100)\\ 		 
		 Week 20 CD4     & 408 (396-421)   & 408 (396-421) & 409 (396-423)\\
		 \hline
	\end{tabular}
\label{Tab:GridSearch}
\end{table}
As before, we compare the performance of the Int$ \mathcal{GP} $, HM$ \mathcal{GP} $, and HE$ \mathcal{GP} $. The analysis presented makes use of the $Mat\acute{e}rn_{5/2}$ covariance, and in Appendix F we present the result for a $Mat\acute{e}rn_{3/2}$ covariance. A natural initial number of design points comes from creating a grid in increments of $ 15 \; kg $ and $ 125 \; cells/\mu L $, from $ 50 \; kg $ to $ 100 \;kg $,  and from $200\; cells/\mu L$ to  $600 \; cells/\mu L $, respectively. This yields a total of 16 design points which is of the order explored by \citep{loeppky2009choosing}. We investigated sampling up to an additional 25 points. By this point, the Expected Improvement relating to the HM$ \mathcal{GP} $ and HE$\mathcal{GP}$ had reached a plateau and $ \psi_W^{opt} $, $ \psi_{CD4}^{opt} $ had converged around a point; the Int$\mathcal{GP}$ did not yet show signs of complete convergence at 25 additional points, but this is not surprising, as we know in a noisy optimization setting, the interpolating method is not the most appropriate approach.  Figure \ref{CaseStudyResulties} shows the estimated value function for each of the modeling approaches . All three yield approximately the same conclusion regarding where the optimal regime may be, and all three methods have focused on choosing additional points in the high $ \psi_w $ region. As we noted previously, the IPW-surface is only moderately noisy. Consequently, it is not surprising that the resulting curves appear to yield similar inference, even with the interpolating $ \mathcal{GP} $ model. When a noisier estimator is used, like the un-normalized IPW estimator, we have observed the interpolating model to yield an estimated response surface that is flat everywhere except for regions very near already sampled points. This emphasizes the fact that although the proposed methodology can be used with any off-the-shelf estimator, the resulting inference can be impacted by the properties of the chosen estimator. 
As with the grid-search approach, in addition to estimating the optimal regime, we are  interested in providing a measure of uncertainty about this optimum. Again, there are two sources of uncertainty to consider. The first is the sampling uncertainty: how will the estimated optimum change across samples. The second relates to the uncertainty represented in the posterior $ \mathcal{GP} $. This posterior informs us about uncertainty in the functional relationship between inputs and outputs, having explored  a finite number of points in index set. Consequently this can also inform us about the uncertainty in the maximizer of the functional relationship between inputs and output. To further characterize what this uncertainty represents, we should consider that if we were to sample the index space very densely; the posterior uncertainty around the curve and the consequent maximizer would be minimized. However, densely sampling the index space does not mean that the uncertainty in the maximizer has gone to zero, as there still remains sampling uncertainty.  

We first examine how to quantify the posterior uncertainty and we then explore how to incorporate sampling uncertainty. To obtain a measure of the posterior uncertainty in the maximizer, after having explored an additional $+m $ points, we first compute the model parameters at $+m$ observations, and we then draw $ N $ sample paths from the posterior. For each sample path, we compute the optimizer to obtain a distribution for the optimal regime. Then, to incorporate the sampling uncertainty, we can use the Bayesian bootstrap for this procedure over 500 replicates. Ultimately, this yields a distribution of optimal regimes that represent both sources of uncertainty. We have implemented this for all three $ \mathcal{GP} $ modeling approaches and the results are presented in Table \ref{TabResulties}. We note that the HM$\mathcal{GP}$ and HE$\mathcal{GP}$ arrive at very similar conclusions, after having explored $ +25 $ points, and that the median optimal regime in the interpolating approach is in the credible interval of the other two methods.  We note additionally that the credible intervals of the Int$\mathcal{GP}$ are much wider than those of the other two methods; the HE$\mathcal{GP}$ approach results in slightly tighter credible intervals for the $ \psi^{CD4}_{opt} $ parameter. Inference at $ +15 $ points is very similar to that which results at $ +25 $ points. The estimated optimal regime is at thresholds of $ 98 \; kg $ and $ 290 \; cells/\mu L $, yielding an expected CD4 cell count of $ 408  \; cells/\mu L $. There is considerable uncertainty regarding the optimal threshold in the $\psi_{CD4}$ direction, but this can be understood from the relatively flat relationship that appears for high values of $ \psi_W $, as can be well explored in the \href{https://danroduq.github.io/ObjectiveII/case-study.html#normalized-ipw-surface}{Interactive Supplement}. Producing Table \ref{TabResulties} is a computationally intensive procedure. In Appendix F, we discuss some efficiency considerations for performing this analysis.

\begin{table}[H]
	\centering
	\caption{HIV Study: Estimates and 95 \% credible intervals for each $ \mathcal{GP} $ modeling strategy; 250 sample paths; 500 Bayesian bootstrapped samples.} 
	\begin{tabular}{rlllll}
		\hline
		Model & Parameter & +1 & +5 & +15 & +25 \\ 
		\hline
		Int$ \mathcal{GP} $   & $ \hat{\psi}^W_{opt} $      & 98 (66-98)            & 98 (58-98)             & 94 (54-98)          & 94 (54-98) \\ 
		Int$ \mathcal{GP} $   & $ \hat{\psi}^{CD4}_{opt} $  & 365 (200-597.5)       & 305 (200-597.5)        & 327.5 (200-597.5)   & 357.5 (200-597.5) \\ 
		Int$ \mathcal{GP} $   & 20 Week CD4     & 409.2 (397.1-421.7)   & 408.7 (397.2-421.09)   & 409.4 (398.0-425.5) & 410.1 (398.3-426.3) \\ 
		\hline
		HM$ \mathcal{GP} $   & $ \hat{\psi}^W_{opt} $     & 98 (66-98)             & 98 (66-98)             & 98 (78-98)             & 98 (78-98) \\ 
		HM$ \mathcal{GP} $   & $ \hat{\psi}^{CD4}_{opt} $ & 357.5 (200-597.5)      & 305 (200-597.5)        & 290 (200-522.5)        & 290 (200-492.5) \\ 
		HM$ \mathcal{GP} $   & 20 Week CD4    & 409.0 (397.0-421.4)    & 408.3 (396.8-420.5)    & 408.3 (397.15-420.2)   & 408.2 (397.3-420.5) \\ 
		\hline
		HE$ \mathcal{GP} $ & $ \hat{\psi}^W_{opt} $      & 98 (70-98)           & 98 (66-98)            & 98 (74-98)             & 98 (78-98) \\ 
		HE$ \mathcal{GP} $ & $ \hat{\psi}^{CD4}_{opt} $  & 350 (200-597.5)      & 305 (200-597.5)       & 290 (200-515)          & 290 (200-462.5) \\ 
		HE$ \mathcal{GP} $ & 20 Week CD4     & 408.9 (397.0-421.4)  & 408.2 (396.6-420.4)   & 408.1 (396.8-420.3)    & 408.4 (396.9-420.5) \\ 
		\hline
	\end{tabular}
	\caption*{\small Increments for the sample paths were by $ 4 kg $ in the $ \psi_W $ axis and by $ 7.5 \; cells/\mu L $ in the $ \psi_{CD4}$ axis}
	\label{TabResulties}
\end{table}

The purpose of this case study was to show how an off-the-shelf estimator could be combined with $ \mathcal{GP} $ techniques in order to arrive at a conclusion about the optimal regime. We saw that the homoskedastic and heteroskedastic analyses produce similar inference. Overall, we can conclude that there are regions of higher and lower value in the value function and that  based on the HM$ \mathcal{GP} $ and HE$ \mathcal{GP} $ there is an optimal threshold of $ 98 \; gk $ and $ 290 \; cells/\mu L $ in the weight and CD4 direction, respectively. There  is relatively low uncertainty around $ \psi^W_{opt} $, but there still remains high levels of uncertainty around $ \psi^{CD4}_{opt} $.

\section{Discussion}

We have been motivated by the possibility that some value-search estimators may not be robust in identifying optimal DTRs, in particular Dynamic MSMs or a grid-search. We explored whether a Bayesian optimization approach via $ \mathcal{GP}s $ could allow for the inference of optimal DTRs. We determined that the estimation surface resulting from the use of an estimator for the value of a DTR tends to exhibit a non-smooth quality resulting from the point-wise variation of the estimator. This led us to examine possible sources of variability in the estimation surface and to consider approaches that allow for varying noise structures. Via simulation studies, we examined the performance of three $ \mathcal{GP} $ methods and found that out of these methods the HM$ \mathcal{GP} $ and HE$ \mathcal{GP} $ consistently yielded comparable or more accurate and precise inference than the Int$ \mathcal{GP} $. Simulations also showed that a grid-search is not always the most accurate approach with $ \mathcal{GP} $ methods tending to provide more accurate and precise results. These methods also required significantly less estimator evaluations to arrive at an estimate for the optimum, thereby making them more efficient than a grid-search. We conclude that there can be much to gain in using an HM$ \mathcal{GP} $ or HE$ \mathcal{GP} $. The performance of the HM$ \mathcal{GP} $ and HE$ \mathcal{GP} $ was similar across all twelve simulations, except in simulation II, where for a sample size of $ n=500 $, the HE$ \mathcal{GP} $ yielded more precise inference slightly faster than the HM$ \mathcal{GP} $. After a few extra sampled points, the HM$ \mathcal{GP} $ achieved comparable inference to the HE$ \mathcal{GP} $ and that the HM$ \mathcal{GP} $ is much more computationally efficient to fit, and therefore we would recommend utilizing the HM$\mathcal{GP}$ in general applications. The comparable inference that these two methods yield was confirmed in the case study, which additionally served to emphasized that this methodology can be applied meaningfully in order to identify an optimal decision rule. Additionally, the case study allowed us to examine how both sampling and posterior uncertainty in the value function can be well represented. Future work should look to examine whether a fully Bayesian treatment benefits the inferential process. Additionally, examining whether the $ \mathcal{GP} $ can jointly represent sampling variability in addition to uncertainty about the value function is an important area of investigation.  Studying the use of other infill criteria and examining the consequences of different stopping rules is also of methodological interest.

\bibliographystyle{abbrvnat}
\bibliography{Paper2_Bibliography_Fixed}

\end{document}